\input ./style/arxiv-general.cfg
\documentclass[aap,MSNbibl,nameyear,dvips]{arximspdf}
\makeatletter
   \@ifpackageloaded{graphicx}{}{\usepackage{graphicx}}
\makeatother
\doi{10.1214/14-AAP1043} %kopijuoti is PTS
\volume{25}
\issue{4}
\pubyear{2015}
\firstpage{2066}
\lastpage{2095}
\docsubty{FLA}

\makeatletter
\newcommand{\rrvert}{\vert}
\newcommand{\llvert}{\vert}
\newcommand{\eqref}[1]{(\ref{#1})}
\newtheorem{theorem}{Theorem}[section]
\newtheorem{proposition}[theorem]{Proposition}
\newtheorem{lemma}[theorem]{Lemma}
\newtheorem{corollary}[theorem]{Corollary}
\newproclaim{remark}[theorem]{Remark}
\newproclaim{definition}[theorem]{Definition}
\newtheorem{example}[theorem]{Example}
\newproclaim{assumption}[theorem]{Assumption}

\newcommand\reals{\mathbb R}
\makeatother

\begin{document}
\begin{frontmatter}

%\dochead{}
\title{Hedging, arbitrage and optimality with superlinear~frictions}
\runtitle{Superlinear frictions}

\begin{aug}
\author[A]{\fnms{Paolo}~\snm{Guasoni}\thanksref{T2}\ead[label=e1]{guasoni@bu.edu}}
\and
\author[B]{\fnms{Mikl\'os} \snm{R\'asonyi}\corref{}\ead[label=e2]{rasonyi.miklos@renyi.mta.hu}}
%
%\author{\fnms{}~\snm{}\corref{}}
\runauthor{P. Guasoni and M. R\'asonyi}
\affiliation{Boston University and Dublin City University,
and MTA Alfr\'ed R\'enyi Institute of Mathematics,
Budapest and University of Edinburgh}

\address[A]{Department of Mathematics and Statistics\\
Boston University\\
111 Cummington Street\\
Boston, Massachusetts 02215\\
USA\\
and\\
School of Mathematical Sciences\\
Dublin City University\\
Glasnevin, Dublin 9\\
Ireland\\
\printead{e1}}
\address[B]{MTA Alfr\'ed R\'enyi Mathematical Institute\\
Re\'altanoda utca 13-15\\
1053 Budapest\\
Hungary\\
and\\
School of Mathematics\\
University of Edinburgh\\
King's Buildings\\
Edinburgh, EH9 3JZ\\
United Kingdom\\
\printead{e2}}
%\address{Address of the Third author,\\\printead{e3,u1}}

%\and
%\author{\fnms{}~\snm{}}
%\runauthor{}
%\affiliation{}
%\dedicated{}
%\address{} %adresu isvedimo komanda gale!
%\address{}
\end{aug}

% HISTORY:
\received{\smonth{8} \syear{2013}}
\revised{\smonth{3} \syear{2014}}
%\accepted{\smonth{} \syear{}}

% ABSTRACT
%
\begin{abstract}
In a continuous-time model with multiple assets described by c\`adl\`ag
processes, this paper characterizes superhedging prices, absence of
arbitrage, and utility maximizing strategies, under general frictions
that make execution prices arbitrarily unfavorable for high trading intensity.
Such frictions induce a duality between feasible trading strategies and
shadow execution prices with a martingale measure. Utility maximizing
strategies exist even if arbitrage is present, because it is not
scalable at will.
\end{abstract}

% KEYWORDS
% Pirmas kwd is didziosios raides
%

\thankstext{T2}{Supported in part by the ERC (278295),
NSF (DMS-11-09047),
SFI (07/SK/M1189, 08/SRC/FMC1389), and FP7 (RG-248896).}
%\begin{keyword}[class=AMS]
%\kwd[Primary ]{}
%\kwd{}
%\kwd[; secondary ]{}
%\end{keyword}
%\begin{keyword}
%\kwd{}
%\end{keyword}
\end{frontmatter}
%
%s1 #&#
\section{Introduction}\label{sec1}

In financial markets, trading moves prices against the trader: buying
faster increases execution prices, and selling faster decreases them.
This aspect of liquidity, known as market depth [\citet{black1986noise}]
or price-impact, is widely documented empirically [\citet
{dufour2000time,rama}], and has received increasing attention in models
of asymmetric information [\citet{kyle}], illiquid portfolio choice
[\citet
{rogers-singh,garleanu-pedersen}] and optimal liquidation [\citet
{almgren-chriss,bertsimas-lo,schied-schoneborn}].
These models depart from the literature on frictionless markets, where
prices are the same for any amount traded. They also depart from
proportional transaction costs models, in which prices differ for
buying and selling, but are insensitive to quantities.\setcounter{footnote}{1}\footnote{A
separate class of models [e.g., \citeauthor{bank2011model}
(\citeyear{bank2011model,bank2011model2})]
investigates the conditions under which current prices depend on past
trading activity, a distinct effect also referred to as (permanent)
price-impact. This paper focuses on temporary price-impact, or market
depth, which some authors call nonlinear transaction costs [cf. \citet{garleanu-pedersen}].}

The growing interest in price-impact has also highlighted a shortage of
effective theoretical tools. In these models, what is the analogue of a
martingale measure? Which contingent claims are hedgeable, and at what
price? How do the familiar optimality conditions for utility
maximization look in this context?
In discrete time, several researchers have studied these fundamental
questions [\citet{astic-touzi,pennanen-penner,pennanen,ds}], but
extensions to continuous time have proved challenging. This paper aims
at filling the gap.

%While in discrete time price-impact is typically a function of traded
%quantities \citet{jarrow1992market}, the same approach in continuous
%time leads to vanishing price-impact
%\citet{cetin2004liquidity,cetin2007option,cetin2007modeling}, and needs
%to be replaced by \footnote{\citet{cetin2007option} obtain nontrivial
%effects using constraints the set of trading strategies, although
%their economic interpretation remains an open question.}

Tackling price-impact in continuous-time requires to clarify two basic
concepts that remain concealed in discrete models:
the relevant classes of trading strategies and of dual variables.
First, to retain price-impact effects in continuous time, execution
prices must depend on the traded quantities per unit of time, that is,
on trading intensity, rather than on the traded quantities themselves,
otherwise price-impact can be avoided with judicious policies [\citet
{cetin2004liquidity,cetin2007option,cetin2007modeling}].
Various classes of trading strategies have appeared in different models
[\citet{cetin2007option,schied-schoneborn}], but a generally agreed
definition of what kind of strategies should be allowed has not yet
emerged. The second key concept is the relevant notion of dual
variables---the analogue of a martingale measure.
The proportional transaction costs literature identifies the
corresponding dual variable as a
consistent prices system, a pair $(\tilde S, Q)$ of a price $\tilde S$
evolving within the bid-ask spread, and a
probability $Q$ under which $\tilde S$ is a martingale.\footnote{These dual objects first
appeared in \citet{jouini}. They were baptized ``consistent price
systems'' in \citet{walter}. See
\citet{MR2589621} for further developments.} Such a definition suggests
that with frictions, passing
to the risk-neutral setting requires both a change in the probability
measure and a change in the price process.

%Such a definition is not directly applicable to price-impact models,
%since bid-ask spreads are typically absent.

%Second, the concept of instantaneous portfolio value is intrinsically
%multivariate, because immediate liquidation is impossible when trading
%can take place only at finite rates, as in price-impact models.

Superlinear frictions in the sense of the present paper, such as
price-impact models,
entail that execution prices become arbitrarily unfavorable as traded
quantities per unit of time grow:
buying or selling too fast becomes impossible. As a result, trading is
feasible only at finite rates---the number of shares is absolutely continuous. This feature sets apart
superlinear frictions from
frictionless markets, in which the number of shares is merely
predictable, and from models with proportional transaction costs, in
which they have finite variation.

Finite trading rates have two central implications: first, portfolio
values are well defined for asset prices
that follow general c\`adl\`ag processes,
not only for semi-martingales. Second, immediate portfolio liquidation
is impossible and, therefore, the
usual notion of admissibility,
based on a lower bound
for liquidation values, is inappropriate. We define below a \emph
{feasible} strategy as any
trading policy with finite trading rate and trading volume, without any
lower bounds on portfolio values.
In particular, this definition does not involve the asset price. In
frictionless markets,
or under proportional transaction costs, this approach would fail for
two reasons: first,
the set of claims attainable by feasible strategies would not be closed
in any reasonable sense,
as a block trade is approximated by
intense trading over small time intervals. Second, portfolios unbounded from
below allow doubling strategies, which lead to arbitrage even with
martingale prices.

Neither issue arises in our models with superlinear frictions.
Block trades are infeasible, even in the limit, as intense trading
incurs exorbitant costs:
put differently, bounded losses imply bounded \emph{trading volume}
(Lemma~\ref{morecamb}).
The bound on trading volume in turn yields the closedness of the
payoffs of feasible strategies
(Proposition~\ref{key}), and the martingale property of portfolio
values under shadow execution prices,
which excludes arbitrage through doubling strategies (Lemma~\ref{martingale}).

Arbitrage also occurs differently in the present setting. Unlike models
without friction or with proportional transaction costs, where an
arbitrage opportunity scales freely, superlinear frictions imply that
scaling trading rates results in a less than proportional scaling of
payoffs [see \citet{deflator} for more about scalable arbitrage].
In fact, in our setting (Assumption~\ref{below}) we prove a stronger
result, whereby \emph{all} payoffs are dominated
by a single random variable, the \emph{market bound}, which depends on
the friction and on the asset price only (Lemma~\ref{key}). This bound
implies that price-impact defeats arbitrage, if pursued on a large scale.

All these definitions and properties come together in the main
superhedging result,
Theorem~\ref{superreplication2}, which characterizes the initial asset
positions that
can dominate a given claim through trading, in terms of shadow
execution prices.
The main message of this theorem is that the superhedging price of a
claim is the supremum of its
expected value under a martingale measure for a shadow execution price,
\emph{minus} a penalty,
which reflects how far the shadow price is from the base price. The
penalty depends on the \emph{dual friction}, introduced by \citet{ds} in
discrete time, and is zero for any equivalent martingale measure of the
asset price. Importantly, the theorem is valid even if there are no
martingale measures, or if the price is not a semi-martingale.

The superhedging theorem, which does not assume absence of arbitrage,
characterizes a large class of models that do not admit arbitrage of
the second kind (strategies that lead to a sure minimal gain) even in
limited amounts.
As for proportional transaction costs, this class contains any price
process that satisfies the conditional full support property \citet{grs}, including fractional Brownian motion.

We conclude the paper by addressing utility maximization.
First, a general theorem guarantees that optimal solutions exist.
This holds true even in the eventual presence of arbitrage
opportunities, which must be chosen optimally,
lest price-impact offset gains. Second, optimal strategies are
identified by a version of the familiar first-order condition that the
marginal utility of the optimal payoff be proportional to a stochastic
discount factor. Technicalities aside, price-impact leads to a novel
condition, which prescribes that a stochastic discount factor makes the
shadow execution price, not the base price, a martingale.
In models with proportional transaction costs this criterion formally
reduces to the usual shadow price approach for optimality [\citet
{kallsen2010using}].

The rest of the paper proceeds with Section~\ref{222}, which describes
the model in detail. The main theoretical tools are developed in
Section~\ref{333}, which proves the market bound, the trading volume
bound, the closedness of the payoff space, and the main superhedging
result. Section~\ref{444} discusses the implications for arbitrage of
the second kind, and its absence with prices with conditional full
support. Section~\ref{555} concludes with the results on utility maximization.

%s2 #&#
\section{The model}\label{222}

For a finite time horizon $T>0$, consider a filtered probability space
$(\Omega,\mathcal{F},(\mathcal{F}_t)_{t\in[0,T]},P)$ with $\mathcal
{F}_0$ trivial, satisfying the usual hypotheses as well as $\mathcal
{F}=\mathcal{F}_T$. $\mathcal{O}$ denotes the optional sigma-field on
$\Omega\times[0,T]$.
The market includes a riskless and perfectly liquid asset $S^0$, used
as numeraire, hence $S^0_t\equiv1$, $t\in[0,T]$, and $d$ risky
assets, described by c\`adl\`ag, adapted
processes $(S^i_t)^{1\le i\le d}_{t\in[0,T]}$. Henceforth, $S$ denotes
the $d$-dimensional process with components $S^i$, $1\le i\le d$, the
concatenation $xy$ of two vectors $x,y$ of equal dimensions denotes
their scalar product, and $|x|$ denotes the Euclidean norm of $x$. The
components of a $(d+1)$-dimensional vector $x$ are denoted by
$x^0,\ldots,x^d$.
%we will denote by $\hat{x}$ the $d$-dimensional vector formed from its
%last $d$ coordinates.

%At time $t$, the investor holds $\Phi_t^i$ shares of the $i$-th asset.
The next definition identifies those strategies for which the number of
shares changes over time at some finite rate $\phi$, hence it is
absolutely continuous.

%de2.1 #&#
\begin{definition}\label{def:feasible}
A \emph{feasible strategy} is a process $\phi$ in the class
%
%e1 #&#
\begin{equation}
\mathcal{A}:= \biggl\{\phi\dvtx \phi\mbox{ is a $\reals^d$-valued,
optional process},\int_0^T \vert
\phi_u\vert \,du<\infty\mbox{ a.s.} \biggr\}.
\end{equation}
%
%and $\Phi_t^i:=\Phi_0^i+\int_0^t \phi_u^i du$ denotes the
%corresponding number of shares.
\end{definition}

% $\Phi^{1\le i\le d}_{t\in[0,T]}$ is

In this definition, the process $\phi$ represents the \emph{trading
rate}, that is, the speed at
which the number of shares in each asset changes over time, and the condition
$\int_0^T \vert\phi_u\vert \,du<\infty$ means that \emph{absolute turnover}
(the cumulative number of shares bought or sold) remains finite in
finite time.

The above definition compares to the one of admissible strategies in
frictionless markets as follows.
On one hand, it relaxes the solvency constraint typical of
admissibility, since a feasible strategy can lead to negative wealth.
On the other hand, this definition restricts the number of shares to be
differentiable in time, while usual admissible strategies have an
arbitrarily irregular number of
shares.\footnote{In the definition of feasible strategy an optional
trading rate leads to a continuous, hence predictable, number of
shares, as for usual admissible strategies.} Note also that the definition
of feasibility does not involve the asset price at all.

With this notation, in the absence of frictions the self-financing
condition would imply a position at time $T$ in the safe asset
(henceforth, cash) equal to:\footnote{By the c\`adl\`ag property
of $S_t$, the function $S_t(\omega),t\in[0,T]$ is
bounded for almost every $\omega\in\Omega$, hence the integral in
\eqref
{lastintegral} is finite a.s. for each $\phi$ satisfying $\int_0^T\vert
\phi_t\vert \,dt<\infty$ a.s.}
%
%e2 #&#
\begin{equation}
\label{lastintegral} z^0-\int_0^T
S_t\phi_t \,dt,
\end{equation}
where $z^0$ represents the initial capital, and the integral reflects
the cost of purchases and the proceeds of sales. For a given trading
strategy $\phi$, frictions reduce the cash position, by making
purchases more expensive, and sales less profitable. With a similar
notation to \citet{ds}, we model this effect by introducing a function
$G$, which summarizes the impact of frictions on the execution price at
different trading rates:

%as2.2 #&#
\begin{assumption}[(Friction)]\label{def:frict}
Let $G\dvtx \Omega\times[0,T]\times\mathbb{R}^d\to\mathbb{R}_+$ be a
$\mathcal{O}\otimes\mathcal{B}(\mathbb{R}^d)$-measurable function, such
that $G(\omega,t,\cdot)$ is convex with $G(\omega,t,x)\geq G(\omega,t,0)$ for all $\omega,t,x$. Henceforth, set $G_t(x):=G(\omega,t,x)$,
that is, the dependence on $\omega$ is omitted, and $t$ is used as a subscript.
\end{assumption}

With this definition, for a given strategy $\phi\in\mathcal{A}$ and an
initial asset position $z\in\reals^{d+1}$, the resulting positions at
time $t\in[0,T]$ in the risky and safe assets are defined as
%
%e3 #&#
%e4 #&#
\begin{eqnarray}
\label{eq:w} V^i_t(z,\phi)&:=& z^i+\int
_0^t \phi_u^i \,du, \qquad 1\le
i\le d,
\\
\label{eq:selffin} V^0_t(z,\phi) &:=& z^0-\int
_0^t \phi_u S_u \,du-
\int_0^t G_u(\phi_u)
\,du.
\end{eqnarray}
The first equation merely says that the cumulative number of shares
$V^i_t$ in the $i$th asset is given by the initial number of shares,
plus subsequent flows.
The second equation contains the new term involving the friction $G$,
which summarizes the impact of trading on execution prices. The
condition $G(\omega,t,x)\geq G(\omega,t,0)$ means that inactivity is
always cheaper than any trading activity. Most models in the literature
assume $G(\omega,t,0)=0$, but the above definition allows for
$G(\omega,t,0)>0$, which is interpreted as a cost of participation in the
market, such as the fees charged by exchanges to trading firms, or as a
monitoring cost.
The convexity of $x\mapsto G_t(x)$ implies that, excluding monitoring costs,
trading twice as fast for half the time locally increases execution
costs---speed is
expensive.\footnote{Let $g(x)=G(\omega, t, x)$, that is, focus on a
local effect. Then, by convexity, $g(x) \le(1-1/k)g(0) + (1/k) g(k x)$
for $k>1$ and, therefore, $(g(k x)- g(0))T/k \ge(g(x) - g(0))T$, which
means that increasing trading speed by a factor of $k$ and reducing
trading time by the same factor implies higher trading costs, excluding
the monitoring cost captured by $g(0)$.}
Finally, note that in general $V^0_t$ may take the value $-\infty$ for
some (unwise) strategies.

With a single risky asset and for $G(\omega,t,0)=0$, the above specification
is equivalent to assuming that a trading rate of $\phi_t\neq 0$ implies an
execution price equal to
%
%e5 #&#
\begin{equation}
\label{barnabe} \tilde S_t = S_t + G_t(
\phi_t)/\phi_t,
\end{equation}
which is (since $G$ is positive) higher than $S_t$ when buying, and
lower when selling.
%With multiple risky assets, price impact cannot be separated across
%assets, and the above specification allows to capture the impact of
%trading rates on each asset both on itself and on other assets.
Thus, $G \equiv0$ boils down to a frictionless market, while
proportional transaction costs correspond to $G_t (x) = \varepsilon S_t
|x|$ with some
$\varepsilon>0$. Yet this paper focuses on neither of these settings,
which entail either zero or linear costs, but rather on
superlinear frictions, defined as those that satisfy the following
conditions. Note that we require a strong form of superlinearity here
(i.e., the cost functional grows at least as a superlinear power of the
traded volume).

%as2.3 #&#
\begin{assumption}[(Superlinearity)]\label{below}
There is $\alpha>1$ and an optional process $H$ such that\footnote{We
implicitly assume that $\inf_{t\in[0,T]} H_t$ is a random
variable, which is always the case if, for example, $H$ is c\`adl\`ag.}
%
%e6 #&#
%e7 #&#
%e8 #&#
%e9 #&#
\begin{eqnarray}
\label{eq:hpositive} \inf_{t\in[0,T]} H_t& >& 0\qquad \mbox{a.s.},
\\
\label{eq:superlinear} G_t(x) &\geq& H_t \vert x
\vert^{\alpha}\qquad \mbox{for all }\omega,t,x,
\\
\label{eq:locint} \int_0^T \Bigl(\sup
_{|x|\leq N}G_t(x) \Bigr)\,dt& <& \infty\qquad\mbox{a.s. for all
}N>0,
\\
\label{eq:0b} \sup_{t\in[0,T]}G_t(0)&\leq& K\qquad\mbox{a.s.
for some constant }K.
\end{eqnarray}
\end{assumption}

Condition \eqref{eq:superlinear} is the central superlinearity
assumption, and prescribes that trading twice as fast for half the time
increases trading costs (in excess of monitoring) by a minimum positive
proportion. Condition \eqref{eq:hpositive} requires that frictions
never disappear, and \eqref{eq:locint} that they remain finite in
finite time. By \eqref{eq:0b}, the participation cost must be uniformly
bounded in $\omega\in\Omega$.
In summary, these conditions characterize nontrivial, finite,
superlinear frictions. Note that \eqref{eq:superlinear} implies that
$\tilde{S}_t$ in \eqref{barnabe} becomes arbitrarily negative as
$\phi
_t$ becomes negative enough, that is, when selling too fast. This issue
is addressed in more detail in Remarks \ref{remi} and \ref{remi2} below.

The most common examples in the literature are, with one risky asset,
the friction $G_t(x):=\Lambda\vert x\vert^{\alpha}$ for some
$\Lambda
>0$, $\alpha> 1$ [see, e.g., \citet{ds}] and, in multiasset models,
the friction
$G_t(x):=x' \Lambda x$ for some symmetric, positive-definite, $d\times
d$ square matrix $\Lambda$ (here $x'$ stands for the transposition
of the vector $x$); see \citet{garleanu-pedersen}.

%re2.4 #&#
\begin{remark}
We conjecture that \eqref{eq:superlinear} could be weakened to
the superlinearity condition
\[
\lim_{x\to\infty}G_t(x)/|x|=\infty\qquad\mbox{a.s.,}
\]
using
Orlicz spaces instead of $L^p$-estimates (i.e., H\"older's inequality).
This generalization is expected
to involve substantial further technicalities for a limited increase in
generality, hence it is not pursued here.
\end{remark}

%re2.5 #&#
\begin{remark}
Our results remain valid assuming that \eqref{eq:superlinear}
holds for $\vert x\vert\geq M$ only, with some $M>0$. Such an
extension requires only minor modifications of the proofs, and may
accommodate models for which a low trading rate incurs either zero or
linear costs.
\end{remark}

%\begin{remark} {\rm The case $G_t(x)=\Lambda|x|$ (i.e. $\alpha=1$)
%would correspond to transaction costs proportional to the traded
%volume, which has been studied in great detail elsewhere, see e.g.
%\citet{MR2589621}. Our present study concentrates on the case $
%\alpha>1$, i.e. on superlinear liquidityfunctions. We will see that
%Assumption~\ref{below} enables us to establish a general duality
%theory for superhedging and arbitrage.} \end{remark}
%One could weaken Assumption~\ref{below} to
%\[
%g_t(x)\geq h_t \vert x\vert^{\alpha},
%\]
%for all $\omega,t$ and $\vert x\vert\geq1$ (i.e. only for large
%enough $x$)
%at the price of some extra work. Do you think it is worth pursuing
%this generalization ?
%In fact, it is enough to have $g_t(x)/\vert x\vert\to\infty$, $x\to
%\infty$ for
%all $t,\omega$, but this would require substantially more work.

%Assumption~\ref{bounded} is a very weak form of boundedness. E.g. if
%$g_u(x):=f(S_u,x)$ for some
%locally bounded and measurable $f$ then Assumption~\ref{bounded}
%automatically holds.

%s3 #&#
\section{Superhedging and dual characterization of payoffs}\label{333}

Despite their similarity to models of frictionless markets and
proportional transaction costs, superlinear frictions in the sense of
Assumption~\ref{below} lead to a surprisingly different structure of
attainable payoffs, as shown in this section. Indeed, the class of
feasible strategies considered above, while still well defined even in
a model without frictions or with proportional transaction costs, is
virtually useless in such settings, as the set of terminal payoffs
corresponding to feasible strategies is not closed in any reasonable sense.

As an example, a simple trading policy that buys one share of the risky
asset at time $t$ and sells it at time $T$ is not a feasible strategy
in the above sense, because it is not absolutely continuous, and in
fact is discontinuous at $t$ and $T$. Yet, in frictionless markets or
with transaction costs, this policy is approximated arbitrarily well by
another one that buys at rate~$n$ in the interval $[t,t+1/n]$ and sells
at rate $n$ on $[T,T+1/n]$. That is, the sequence of corresponding
payoffs converges to a finite payoff, but this limit payoff does not
belong to the payoff space of feasible strategies.

By contrast, with the superlinear frictions in Assumption~\ref{below},
the set of terminal values corresponding to feasible strategies \emph
{is} closed in a strong sense.
The intuitive reason is that approximating a nonsmooth
strategy would require trading at increasingly high speed, generating
infinite costs, and preventing convergence to a finite payoff.
%Incidentally, infinite turnover would also cause high losses, but we
%have relaxed the traditional admissibility, therefore this aspect is
%irrelevant for feasible strategies.

%s3.1 #&#
\subsection{The market bound}
Superlinear frictions in the sense of Assumption~\ref{below} lead to a
striking boundedness property: for a fixed initial position, all
payoffs of feasible strategies are bounded above by a single random
variable $B<\infty$, the \emph{market bound}, which depends on the
friction $G$ and on the price $S$, but \emph{not} on the strategy. This
property clearly fails in frictionless markets, where any payoff with
zero initial capital can be scaled arbitrarily and, therefore, admits
no uniform bound. In such markets, a much weaker boundedness property
holds: Corollary~9.3 of \citet{dsc} shows that the set of payoff of
$x$-admissible strategies is bounded in $L^0$ \emph{if} the market is
arbitrage-free in the sense of the condition (NFLVR), and a similar
result holds with proportional transaction costs under the (RNFLVR)
property [\citet{glr}].

A central tool in this analysis is the function $G^*$, the
Fenchel--Legendre conjugate of $G$, which we call \emph{dual friction}.
Its importance was first recognized by \citet{ds}, who used it to derive
a superhedging result in discrete time. $G^*$ is defined as\footnote
{Note that the supremum can be taken over $\mathbb{Q}^d$, hence $G^*$
is $\mathcal{O}\otimes\mathcal{B}(\mathbb{R}^d)$-measurable. Note also
that, under Assumption~\ref{below}, $G^*_t(\cdot)$ is a finite, convex function satisfying
$G_t^*(x)\geq-K$ for all $x$, see the proof of Lemma~\ref{b}.}
%
%e10 #&#
\begin{equation}
G_t^*(y):= \sup_{x\in\mathbb{R}^d} \bigl(xy-G_t(x)
\bigr), \qquad y\in\mathbb {R}^d, t\in[0,T],
\end{equation}
and the typical case $d=1$, $G_t(x) = \Lambda|x|^{\alpha}$ leads to
$G^*_t(y) =
\frac{\alpha-1}{\alpha} \alpha^{{1}/{(1-\alpha)}}\times  \Lambda^{
{1}/{(1-\alpha)}} |y|^{{\alpha}/{(\alpha-1)}}$ [in particular,
$G^*_t(y) = y^2/(4\Lambda)$ for $\alpha=2$].
The key observation is the following.

%le3.1 #&#
\begin{lemma}\label{finite} Under Assumption~\ref{below}, any $\phi
\in
\mathcal A$ satisfies
\[
V^0_T(z,\phi)\leq z^0+\int
_0^T G^*_t(-S_t)\,dt<
\infty \qquad\mbox{a.s.}
\]
\end{lemma}

\begin{pf}
Indeed, this follows from \eqref{eq:selffin}, the definition of
$G_t^*$, and Lemma~\ref{b} below.
\end{pf}

%le3.2 #&#
\begin{lemma}\label{b}
Under Assumption~\ref{below}, the random variable $B:=\int_0^T
G^*_t(-S_t)\,dt$ is finite almost surely.
\end{lemma}

\begin{pf}
Consider first the case $d=1$. Then, by direct calculation,
%
%e11 #&#
\begin{equation}
\label{eq:star} G^*_t(y)\leq \sup_{r\in\mathbb{R}}
\bigl(ry-H_t|r|^{\alpha} \bigr) = \frac
{\alpha
-1}{\alpha}
\alpha^{{1}/{(1-\alpha)}}H_t^{{1}/{(1-\alpha)}} |y|
^{{\alpha}/{(\alpha-1)}}.
\end{equation}
Noting that $\sup_{t\in[0,T]} \vert S_t\vert$ is finite a.s. by the
c\`
adl\`ag property of $S$, and knowing that $\inf_{t\in[0,T]} H_t$ is a
positive random variable, it follows that
\[
\sup_{t\in[0,T]} G_t^*(-S_t)<\infty\qquad\mbox{a.s.},
\]
which clearly implies the statement. If $d>1$, then note that
%
%e12 #&#
\begin{eqnarray}
G^*_t(y)\leq\sup_{r\in\mathbb{R}^d} \Biggl(\sum
_{i=1}^d r^iy^i-H_t|r|^{\alpha}
\Biggr)&\leq& \sum_{i=1}^d \sup
_{r\in\mathbb{R}^d} \bigl(r^iy^i-(H_t/d)|r|^{\alpha
}
\bigr)
\nonumber
\\[-8pt]
\\[-8pt]
\nonumber
&\leq& \sum_{i=1}^d \sup
_{x\in\mathbb{R}} \bigl(xy^i-(H_t/d)|x|^{\alpha
}
\bigr)
\end{eqnarray}
and the conclusion follows from the scalar case.
\end{pf}

Since $B<\infty$ a.s., it is impossible to achieve a scalable arbitrage:
though a trading strategy may realize an a.s. positive terminal value,
one cannot get an arbitrarily large profit by scaling the trading
strategy (i.e., by multiplying it with large positive constants) since
bigger trading values also enlarge costs.
%, that is an arbitrarily large profit from zero initial capital with
%positive probability.
Even if an arbitrage exists, amplifying it too much backfires, because
the superlinear friction eventually overrides profits. Yet, arbitrage
opportunities can exist in limited size (cf. Section~\ref{444} below).

Limited arbitrage opportunities also appear in the frictionless models
of \citet{MR2210925} and \citet{MR2335830} through a completely different
mechanism. These models allow for arbitrage opportunities that can lead
to a possible intermediate loss before realizing a certain final gain,
while requiring that wealth remains positive at all times. As a result,
an arbitrage opportunity is scalable only insofar as its maximal
intermediate loss is less than the initial capital committed to the
arbitrage. By contrast, with superlinear frictions arbitrage is limited
even though wealth may well become negative before gains are realized
(cf. Definition~\ref{def:feasible}), because the superlinear friction
defeats attempts to scale an arbitrage linearly, by reducing and
eventually eliminating its profitability for larger positions.

%s3.2 #&#
\subsection{Trading volume bound}
For $Q\sim P$, denote by $L^1(Q)$ the usual Banach space of
$(d+1)$-dimensional, $Q$-integrable random variables;
given a subset $A$ of a Euclidean space, $L^0(A)$ denotes the set of
($P$-a.s. equivalence classes of)
$A$-valued random variables, equipped with the topology of convergence
in probability.
%$L^0_+$ stands for
%the set of $(d+1)$-dimensional nonnegative random variables.
$E_Q X$ denotes the expectation of a random variable $X$ under $Q$.
From now on, fix $1<\beta<\alpha$, where $\alpha$ is as in Assumption~\ref{below}.
Let $\gamma$ be the conjugate number of $\beta$, defined by
\[
\frac{1}{\beta} +\frac{1}{\gamma}=1.
\]

The next definition identifies a class of reference probability
measures with integrability properties that fit the friction $G$
and the price process $S$ well. Our main results (see Section~\ref{manna}) involve suprema of expectations of various functionals under
families of probability measures equivalent to $P$. Ideally, \emph{all}
such measures should be taken (as in Theorem~\ref{superreplication5}
below) but on infinite $\Omega$ this leads to integrability issues.
Thus, we need to single out a family of probability measures which is
large enough for the results of Section~\ref{manna} to hold, but
also small enough to ensure appropriate integrability properties. This
is why we introduce the sets $\mathcal{P}$ and $\mathcal{P}(W)$ in
Definition~\ref{aksa} below.
$\mathcal{P}$ identifies a set of probabilities under which some shadow
execution price has the martingale property, as explained in the proof
of Theorem~\ref{bagdad} and Lemma~\ref{martingale} below.

%de3.3 #&#
\begin{definition}\label{aksa}
$\mathcal{P}$ denotes the set of probabilities $Q\sim P$ such that
\[
E_Q \int_0^T H_t^{\beta/(\beta-\alpha)}\bigl(1+|S_t|\bigr)^{\beta\alpha
/(\alpha
-\beta)}
\,dt <\infty.
\]
$\tilde{\mathcal{P}}$ denotes the set of probability measures $Q\in
\mathcal{P}$ such that
\[
E_Q \int_0^T |S_t|
\,dt < \infty \quad\mbox{and}\quad E_Q \int_0^T
\sup_{|x|\leq N}G_t(x) \,dt < \infty \qquad\mbox{for all }N
\ge1.
\]
For a (possibly multivariate) random variable $W$, define
\[
\mathcal{P}(W):=\bigl\{Q\in\mathcal{P}\dvtx E_Q|W|<\infty\bigr\},\qquad
\tilde {
\mathcal {P}}(W):=\bigl\{Q\in\tilde{\mathcal{P}}\dvtx E_Q|W|<\infty\bigr\}.
\]
\end{definition}

Under Assumption~\ref{below}, note that $\tilde{\mathcal{P}}(W)\neq
\varnothing$ for all $W$ by
Dellacherie and Meyer [(\citeyear{dm2}), page~266].
The next lemma shows that, if a payoff has a finite negative part under
some probability in $ \mathcal{P}$, then its trading rate must also be
(suitably) integrable. There is no analogue to such a result in
frictionless markets, but transaction costs \citet{glr}, Lemma~5.5, lead
to a similar property, whereby any admissible strategy must satisfy an
upper bound on its total variation. In both cases, the intuition is
that, with frictions, excessive trading causes unbounded losses. Hence,
a bound on losses translates into one for trading volume. Lemma~\ref
{morecamb} is crucial to establish the closedness of the set of
attainable payoffs (Proposition~\ref{key} below) as well as to prove
the martingale property of shadow execution prices in utility
maximization problems (see Lemma~\ref{martingale} in Section~\ref{555}).

In the sequel, $x_-$ denotes the negative part of $x\in\mathbb{R}$.

%le3.4 #&#
\begin{lemma}\label{morecamb}
Let $Q\in\mathcal{P}$ and $\phi\in\mathcal{A}$ be such that
$E_Q\xi
_-<\infty$, where
\[
\xi:=-\int_0^T S_t
\phi_t \,dt-\int_0^T
G_t(\phi_t)\,dt.
\]
Then
\[
E_Q\int_0^T |
\phi_t|^{\beta}\bigl(1+|S_t|\bigr)^{\beta}\,dt<\infty.
\]
\end{lemma}

\begin{pf}
For ease of notation, set $T:=1$. Define $\phi_t(n):=\phi_t 1_{\{
|\phi
_t|\leq n\}}\in\mathcal{A}$, $n\in\mathbb{N}$. As $n\to\infty$, clearly
$\phi_t(n)\to\phi_t$ for all $t$ and $\omega\in\Omega$, and the
random variables
%
%e13 #&#
%e14 #&#
%e15 #&#
\begin{eqnarray}
\xi_n &:= &- \int_0^1
S_t\phi_t(n)\,dt-\int_0^1
G_t\bigl(\phi_t(n)\bigr)\,dt
\\
&=&-\sum_{i=1}^d \int_0^1
S^i_t\phi^i_t(n)
[1_{\{S^i_t\leq0,\phi^i_t\leq0\}}+1_{\{S^i_t> 0,\phi^i_t\leq0\}
}
\nonumber
\\[-8pt]
\\[-8pt]
\nonumber
&&\hspace*{79pt}{}+1_{\{S^i_t\leq0,\phi^i_t> 0\}}+1_{\{S^i_t> 0,\phi^i_t >0\}}]\,dt
\\
&&{} -\int_0^1 G_t\bigl(
\phi_t(n)\bigr)\,dt
\end{eqnarray}
converge to $\xi$ a.s. by monotone convergence. [Note that each of the
terms with an indicator converges monotonically, and that $G_t(0)\leq
G_t(x)$ for all $x$.]
%Furthermore,
%by $Q\in\mathcal{P}$ both integrals in $\xi_n$ are in $L^1(Q)$ and
%$\xi_n\geq\min\{\xi,-K\}$ where $K$ is an upper bound for $\sup_{t\in
%[0,T]}G_t(0)$. Hence, by Fatou's lemma,
%\begin{equation}\label{kep}
%E_Q\xi\leq\liminf_{n\to\infty} E_Q\xi_n.
%\end{equation}
H\"older's inequality yields
%
%e16 #&#
\begin{eqnarray}
\label{krakow} &&\int_0^1 \bigl\vert
\phi_t(n)\bigr\vert^{\beta}\bigl(1+|S_t|\bigr)^{\beta}\,dt\nonumber\\
&&\qquad=
\int_0^1 \bigl\vert\phi_t(n)
\bigr\vert^{\beta}H_t^{\beta/\alpha} \frac{1}{H_t^{\beta/\alpha}}\bigl(1+|S_t|\bigr)^{\beta}
\,dt
\\
\nonumber
&&\qquad\leq\biggl[\int_0^1 \bigl\vert
\phi_t(n)\bigr\vert^{\alpha} H_t \,dt
\biggr]^{\beta/\alpha} \biggl[\int_0^1 \biggl(
\frac{1}{H_t^{\beta/\alpha}}\bigl(1+|S_t|\bigr)^{\beta
} \biggr)^{\alpha/(\alpha-\beta)}\,dt
\biggr]^{(\alpha-\beta)/\alpha}
\\
\nonumber
&&\qquad\leq\biggl[\int_0^1 G_t
\bigl(\phi_t(n)\bigr)\,dt \biggr]^{\beta/\alpha} \biggl[\int
_0^1 \biggl(\frac{1}{H_t^{\beta/\alpha}}\bigl(1+|S_t|\bigr)^{\beta
}
\biggr)^{\alpha/(\alpha-\beta)}\,dt \biggr]^{(\alpha-\beta)/\alpha}.\nonumber
\end{eqnarray}
All these integrals are finite by Assumption~\ref{below} and the c\`
adl\`ag property of $S$.
Now, set
\[
m:= \biggl[\int_0^1 \biggl(
\frac{1}{H_t^{\beta/\alpha
}}\bigl(1+|S_t|\bigr)^{\beta
} \biggr)^{\alpha/(\alpha-\beta)}\,dt
\biggr]^{(\alpha-\beta)/\alpha
},
\]
and note that, by Jensen's inequality,
%
%e17 #&#
\begin{eqnarray}
\label{welch} \biggl\llvert \int_0^1
S_t\phi_t(n) \,dt\biggr\rrvert &\leq&\int
_0^1 \bigl\vert\phi _t(n)\bigr\vert
\bigl(1+|S_t|\bigr) \,dt
\nonumber
\\[-8pt]
\\[-8pt]
\nonumber
&\leq &\biggl[\int_0^1
\bigl\vert\phi_t(n)\bigr\vert^{\beta}\bigl(1+|S_t|\bigr)^{\beta}
\,dt \biggr]^{1/\beta}.
\end{eqnarray}
Note also that if $x\geq1$ and $x\geq2^{{\beta}/{(\alpha-\beta)}}
m^{{\alpha}/{(\alpha-\beta)}}$ then
$x^{1/\beta}-(x/m)^{\alpha/\beta}\leq x-2x=-x$.
This observation, applied to
\[
x:=\int_0^1\bigl \vert\phi_t(n)
\bigr\vert^{\beta}\bigl(1+|S_t|\bigr)^{\beta}\,dt,
\]
implies that $\xi_n\leq-x$ on the event $\{x\geq2^{{\beta
}/{(\alpha
-\beta)}} m^{{\alpha}/{(\alpha-\beta)}}+1\}$.
Thus,
\[
\int_0^1 \bigl\vert\phi_t(n)
\bigr\vert^{\beta}\bigl(1+|S_t|\bigr)^{\beta}\,dt\leq(
\xi_n)_-+ 2^{{\beta}/{(\alpha-\beta)}}
m^{{\alpha}/{(\alpha-\beta)}}+1\qquad \mbox{ a.s.}
\]
Letting $n$ tend to $\infty$, it follows that
%
%e18 #&#
\begin{equation}
\label{lavender} \int_0^1 \vert\phi_t
\vert^{\beta}\bigl(1+|S_t|\bigr)^{\beta}\,dt\leq\xi_-+
2^{
{\beta}/{(\alpha-\beta)}} m^{{\alpha}/{(\alpha-\beta)}}+1,
\end{equation}
which implies the claim, since $E_Q\xi_-<\infty$ by assumption, and
$E_Q m^{{\alpha}/{(\alpha-\beta)}}<\infty$ from $Q\in\mathcal{P}$.
%Hence if we had $\sup_n E_Q\int_0^1 \vert\phi_t(n)\vert^{
%\beta}(1+|S_t|)^{\beta}\,dt=\infty$
%then, along a subsequence,
%\[
%E_Q \xi_n\leq[E_Q\int_0^1 \vert\phi_t(n)\vert^{\beta}(1+|S_t|)^{
%\beta}\,dt]^{1/\beta}-
%[mE_Q\int_0^1 \vert\phi_t(n)\vert^{\beta}(1+|S_t|)^{\beta}\,dt]^{\alpha/
%\beta}\to-\infty,
%\]
%would hold as $n\to\infty$, with
%which is impossible by \eqref{kep} and $E_Q\xi_-<\infty$. The
%statement of the Lemma follows.
%It follows by monotone convergence that
%$$
%\infty>\sup_n E_Q\int_0^1 \vert\phi_t(n)\vert^{\beta}(1+|S_t|)^{
%\beta}\,dt=E_Q\int_0^1 \vert\phi_t\vert^{\beta}(1+|S_t|)^{\beta}\,dt,
%$$
%finishing the proof.
\end{pf}

%s3.3 #&#
\subsection{Closed payoff space}
The central implication of the previous result is that the class of
multivariate payoffs superhedged\vspace*{1pt} by a feasible strategy,
defined as $C:=[\{V_T(0,\phi)\dvtx \phi\in\mathcal{A}\}-L^0(\mathbb
{R}^{d+1}_+)] \cap L^0(\mathbb{R}^{d+1})$, is closed in a rather
strong sense;
recall the componentwise definition of the $(d+1)$-dimensional random
variable $V_T(0,\phi)$
in \eqref{eq:w} and \eqref{eq:selffin}. Closedness is the key
property for
establishing superhedging results; see, for example, Section~9.5 of
\citet{dsc} or Section~3.6 of \citet{MR2589621}.
%hich confirms that this class of strategies is suitable for a
%superhedging result.

%pr3.5 #&#
\begin{proposition}\label{key} Under Assumption~\ref{below},
the set $C\cap L^1(Q)$ is closed in $L^1(Q)$ for all $Q\in\mathcal{P}$
such that $\int_0^T |S_t|\,dt$ is $Q$-integrable.
\end{proposition}

\begin{pf} Take $T=1$ for simplicity, and assume that $\rho_n:=\xi
_n-\eta_n\to\rho$ in $L^1(Q)$ where $\eta_n\in L^0(\mathbb
{R}^{d+1}_+)$ and $\xi_n=V_1(0,\psi(n))$ for some
$\psi(n)\in\mathcal{A}$ are such that
$\rho_n\in L^1(Q)$. Up to a subsequence, this convergence takes place
a.s. as well.

Lemma~\ref{morecamb} implies that $E_Q\int_0^1 \vert\psi_t(n)\vert
^{\beta}(1+|S_t|)^{\beta}\,dt$ must be finite for all $n$
since $(\xi_n)_-\leq(\rho_n)_-$ and the latter is in $L^1(Q)$.
%If we had $\sup_n E_Q\int_0^1 \vert\phi_t(n)\vert^{\beta}(1+|S_t|)^{
%\beta}\,dt=\infty$
Applying \eqref{lavender} with the choice $\phi:=\psi(n)$ yields
\[
\int_0^1 \bigl\vert\psi_t(n)
\bigr\vert^{\beta}\bigl(1+|S_t|\bigr)^{\beta}\,dt\leq(
\rho_n)_-+ 2^{{\beta}/{(\alpha-\beta)}} m^{{\alpha}/{(\alpha-\beta
)}}+1.
\]
Now, since $Q\in\mathcal{P}$, and the sequence $\rho_n$ is bounded in
$L^1(Q)$ because it is convergent in $L^1(Q)$, it follows that
%
%e19 #&#
\begin{equation}
\label{sirius} \sup_{n\ge1} E_Q\int
_0^1 \bigl\vert\psi_t(n)
\bigr\vert^{\beta
}\bigl(1+|S_t|\bigr)^{\beta
}\,dt<\infty.
\end{equation}

%This means that the sequence $\psi_{\cdot}(n)\in L^1(\Omega,
%\mathcal{F},Q;\mathbb{B})$ is norm bounded,
%where $\mathbb{B}:=L^{\beta}([0,T],\mathcal{B}([0,T]),Leb)$ is a
%separable reflexive Banach space and
%$L^1(\Omega,\mathcal{F},Q;\mathbb{B})$ is the Banach space of $
%\mathbb{B}$-valued Bochner integrable functions.
%We apply the infinite dimensional extension of Koml\'os's theorem from
%\citet{balder} and get a subsequence (still
%denoted by $n$) such that $\tilde{\psi}(n):=(1/n)\sum_{j=1}^n \psi(j)$
%(as well as the C\'esaro means of any further
%subsequence) converge to some $\tilde{\psi}_\cdot$ a.s. in the weak
%topology of $\mathbb{B}$.

%would hold as $n\to\infty$,
%which is impossible (note that $\rho_n^0$ denotes the $0$th coordinate
%of $\rho_n$). It follows that
%$\sup_n E_Q\int_0^1 \vert\phi_t(n)\vert^{\beta}(1+|S_t|)^{\beta}\,dt<
%\infty$,

Consider $\mathbb{L}:=L^1(\Omega,\mathcal{F},Q;\mathbb{B})$, the Banach
space of $\mathbb{B}$-valued Bochner-integrable functions, where
$\mathbb{B}:=L^{\beta}([0,1],\mathcal{B}([0,1]),\operatorname{Leb})$ is a separable
and reflexive Banach space. The functions $\psi_{\cdot}(n)\dvtx \Omega\to
\mathbb{B}$ are easily seen to be weakly measurable, hence also
strongly measurable by the separability of $\mathbb{B}$. By \eqref
{sirius}, the sequence $\psi_{\cdot}(n)$ is bounded in $\mathbb{L}$, so
Lemma~15.1.4 in \citet{dsc} yields convex combinations
\[
\tilde{\psi}_{\cdot}(n)=\sum_{j=n}^{M(n)}
\alpha_j(n)\psi_{\cdot}(n),
\]
which converge to some $\tilde{\psi}_{\cdot}\in\mathbb{L}$ a.s. in
$\mathbb{B}$-norm.

By the bound in \eqref{sirius}, $\sup_n E_Q\int_0^1 \vert\phi
_t(n)\vert
(1+|S_t|)\,dt<\infty$.
Now apply Lemma~9.8.1 of \citet{dsc} to the sequence $\tilde{\psi
}_{\cdot}(n)$
in the space of $(d+1)$-dimensional random variables
$L^1(\Omega\times[0,1],\mathcal{O},\nu)$,
where $\nu$ is the measure defined by
\[
\nu(A):= \int_{\Omega\times[0,1]}1_A(\omega,t) \bigl(1+\vert
S_t\vert\bigr) \,dt \,dQ(\omega)
\]
for $A\in\mathcal{O}$ (which is finite by the choice of $Q$).
This lemma yields convex combinations
$\hat{\psi}_{\cdot}(n)$ of the $\tilde{\psi}_{\cdot}(n)$ such
that $\hat
{\psi}_{\cdot}(n)$ converges to $\psi_{\cdot}$ $\nu
$-almost everywhere and hence
$P\times \operatorname{Leb}$-almost everywhere. This shows, in
particular, that ${\psi}$ is $\mathcal{O}$-measurable.

Since $\tilde\psi_\cdot(n)$ converge a.s. in $\mathbb B$-norm, also
$\hat{\psi}_{\cdot}(n)\to\tilde{\psi}$ a.s. in $\mathbb
{B}$-norm, so
$\psi=\tilde{\psi}$, $P\times \operatorname{Leb}$-a.e.
and
hence we may and will assume that $\tilde{\psi}_{\cdot}(n)$ tends
to $\psi$ a.s. in $\mathbb B$-norm as
well as $P \times \operatorname{Leb}$-a.e.

Define $\tilde{\xi}_n:=\sum_{j=n}^{M(n)} \alpha_j(n) \xi_j$ and
$\tilde
{\eta}_n:=\sum_{j=n}^{M(n)} \alpha_j(n) \eta_j$. It holds that\break
$\lim_{n\to\infty}\int_0^1 \tilde{\psi}_t(n)
S_t \,dt=\int_0^1 {\psi}_t S_t \,dt$ almost surely, and also
\[
\lim_{n\to\infty}\tilde{\xi}^i_n=\lim
_{n\to\infty}\int_0^1 \tilde{\psi
}^i_t(n)\,dt=\int_0^1 {
\psi}^i_t \,dt \qquad\mbox{a.s. for }i=1,\ldots,d.
\]
Hence, $\tilde{\eta}_n^i\to\eta^i$ a.s. with $\eta^i:=\int_0^T
\tilde
{\psi}^i_t \,dt-\rho^i\in L^0(\mathbb{R}_+)$. By the convexity of $G_t$,
\begin{eqnarray*}
\rho^0 &=& \lim_{n\to\infty}\bigl(\tilde{
\xi}^0_n-\tilde{\eta}^0_n\bigr)
\\
&\leq&\limsup_{n\to\infty} \biggl[-\int_0^1
\tilde{\psi}_t(n) S_t \,dt-\int_0^1
G_t\bigl(\tilde{\psi}_t(n)\bigr)\,dt- \tilde{
\eta}^0_n \biggr]
\\
&=& \limsup_{n\to\infty} \biggl[-\int_0^1
\tilde{\psi}_t(n) S_t \,dt- \int_0^1
G_t({\psi}_t)\,dt-\int_0^1
G_t\bigl(\tilde{\psi}_t(n)\bigr)\,dt\\
&&\hspace*{173pt}{}+\int
_0^1 G_t({\psi}_t)\,dt-
\tilde{\eta}^0_n \biggr]
\\
&=& -\int_0^1 {\psi}_t
S_t \,dt-\int_0^1 G_t({
\psi}_t)\,dt \\
&&{}+ \limsup_{n\to\infty} \biggl[ -\int
_0^1 G_t\bigl(\tilde{
\psi}_t(n)\bigr)\,dt+ \int_0^1
G_t({\psi}_t)\,dt-\tilde{\eta}_n^0
\biggr].
\end{eqnarray*}
Now Fatou's lemma and $\tilde{\eta}_n\in
L^0(\mathbb{R}^{d+1}_+)$ imply that the limit superior is in
$-L^0(\mathbb{R}_+)$ [note that $G_t(\cdot)$ is continuous
by convexity], hence there is $\eta^0\in L^0(\mathbb{R}_+)$
such that
\[
\rho^0=- \int_0^1 {
\psi}_t S_t \,dt-\int_0^1
G_t({\psi}_t)\,dt -\eta^0,
\]
which proves the proposition.
\end{pf}

%The closedness property above is in fact stronger than closedness in
%probability, as the following corollary shows.

%co3.6 #&#
\begin{corollary}\label{closed}
Under Assumption~\ref{below}, the set $C$ is closed in probability.
\end{corollary}

\begin{pf}
Let $\rho_n\in C$ tend to $\rho$ in probability. Up to a subsequence,
convergence also holds almost surely. There exists $Q\in\mathcal{P}$
[see page 266 of \citet{dm2}]
such that $\rho,\sup_n |\rho-\rho_n|,\int_0^T |S_t|\,dt$ are all
$Q$-integrable.
Then $\rho_n\to\rho$ in $L^1(Q)$ as well, and Proposition~\ref{key}
implies that $\rho\in C$.
\end{pf}

%s3.4 #&#
\subsection{Superhedging}\label{manna}

Finally, the main superhedging theorem.
To the best of our knowledge, Theorem~\ref{superreplication2} is the
first dual characterization in continuous time of hedgeable contingent
claims with price-impact. Results in discrete time include \citet{astic-touzi,pennanen-penner,pennanen,ds}. Our result is inspired, in
particular, by Theorem~3.1 of \citet{ds} for finite probability spaces.

Note that both terminal claims and initial endowments are multivariate,
for a good reason.
Due to the presence of price impact, positions in the safe asset and in
various risky assets are not
immediately convertible into each other at a fixed price.
It is thus impossible to introduce, in a meaningful way, a
one-dimensional wealth process representing holdings in units of a
num\'eraire---multivariate book-keeping of positions is necessary.
%In financial markets In our model price impact is construed as a
%penalization
%of trading fast, in particular, only finite trading speed is permitted.

%With price impact,
%which forces finite trading rates, thereby prohibiting instant
%purchases or sales, even in
%the Black-Scholes model it is impossible to buy one share of the risky
%asset for a sure price in finite time. Thus,
%superhedging of general claims in terms of cash yields mostly trivial
%results.
%Note, however, that the model in \citet{ds} allows liquidation of
%positions at $T$ at price $S_T$ (i.e. without liquidity constraints),
%which leads to a simpler dual characterisation. We think that the
%present framework is more realistic. Under Assumption~\ref{below} it
%is possible to prove the continuous-time version of Theorem~3.1 in
%\citet{ds} along the lines of the proof of Theorem~\ref{superreplication2}.}

In the multivariate notation below, inequalities among vectors are
understood componentwise:
$x\le y$ means that $x^i \le y^i$ for all $i$. Also, for a
$(d+1)$-dimensional vector $x$, define $\bar{x}$ as the
$d$-dimensional vector with $\bar{x}^i=(x^i/x^0)1_{\{x^0\neq0\}}$,
$i=1,\ldots,d$, while $\hat{x}$ denotes the $(d+1)$-dimensional
vector with
components $\hat{x}^i=x^i$, $i=1,\ldots,d$ and $\hat{x}^0=1$. (See
Table~\ref{tab:not} for a summary of notation.)

%t1 #&#
\begin{table}
\caption{Summary of vector notation}
\label{tab:not}
\begin{tabular*}{\textwidth}{@{\extracolsep{\fill}}lcc@{}}
\hline
\multicolumn{1}{@{}l}{$\bolds{\mathbb{ R}}$} &
\multicolumn{1}{c}{$\bolds{\mathbb{ R}^d}$} &
\multicolumn{1}{c@{}}{$\bolds{\mathbb {R}^{d+1}}$}\\
\hline
& $\bar x = (x^1/x^0,\ldots,x^d/x^0)1_{\{x^0\neq0\}}$ & $x =
(x^0,x^1,\ldots,x^d)$\\
& $\tilde x = (x^1,\ldots,x^d)$ & $\hat x = (1,x^1,\ldots,x^d)$\\
$c$ & & $\check c = (c,0,\ldots,0)$\\
\hline
\end{tabular*}
\end{table}

%th3.7 #&#
\begin{theorem}\label{superreplication2}
Let $W\in L^0(\mathbb{R}^{d+1})$, $z\in\mathbb{R}^{d+1}$ and
Assumption~\ref{below} hold.
There exists $\phi\in\mathcal{A}$ such that
$V_T(z,\phi)\geq W$ a.s. if and only if
%
%e20 #&#
\begin{equation}
\label{viola} {Z}_0z\geq E_Q({Z}_TW)-E_Q
\int_0^T Z^0_t
G^*_t(\bar{Z}_t-S_t)\,dt,
\end{equation}
for all $Q\in\mathcal{P}(W)$ and for all $\mathbb{R}^{d+1}_+$-valued
bounded $Q$-martingales $Z$ with $Z_0^0=1$
satisfying $Z^i_t=0$, $i=1,\ldots,d$ on $\{Z^0_t=0\}$.
\end{theorem}

%re3.8 #&#
\begin{remark}\label{remi}
Although the above theorem holds for general $S$, it has the
interpretation of a superreplication
result only if $S$ (or at least $S_T$) has nonnegative components and,
therefore, a positive number of units of risky positions has positive
value. Otherwise, if $S$ can take negative values, a larger number of
units does not imply a position with higher value, but only a larger exposure.

Assume in the rest of this remark that $S$ is nonnegative and
one-dimen\-sional (for simplicity). Take $\phi\in\mathcal{A}$ and consider
the (optional) set $A:=\{(\omega,t)\dvtx \phi_t(\omega)\neq0,
S_t(\omega)
+ G(\omega,t,\phi_t(\omega))/\phi_t(\omega)\geq0\}$, which identifies
the times at which execution prices are positive.
Clearly, $V_T(z,\phi')\geq V_T(z,\phi)$ for\break $\phi_t'(\omega):=\phi
_t(\omega)1_A$. Hence, in Theorem~\ref{superreplication2}
one may replace $\mathcal{A}$ by
\[
\mathcal{A}_+:=\bigl\{\phi\in\mathcal{A}\dvtx S_t(\omega) + G\bigl(
\omega,t,\phi _t(\omega)\bigr)/\phi_t(\omega)\geq0\mbox{
when }\phi_t(\omega)\neq 0\bigr\}.
\]
In other words, the superreplication result continues to hold by
considering only trading strategies with positive execution prices at
all times, because any other strategy is dominated pointwise by a
strategy that trades at the same rate when the execution price is
positive, and otherwise does not trade. The class $\mathcal{A}_+$ is
economically more appealing as it excludes the unintended consequence
of \eqref{eq:superlinear} that $S_t(\omega) + G(\omega,t,\phi
_t(\omega
))/\phi_t(\omega)\to-\infty$ whenever $\phi_t(\omega)\to-\infty$.
\end{remark}

The proof of Theorem~\ref{superreplication2} in fact yields also the
following slightly
different version, in terms of bounded martingales only.

%th3.9 #&#
\begin{theorem}\label{superreplication3}
Let $W\in L^0(\mathbb{R}^{d+1})$, $z\in\mathbb{R}^{d+1}$ and let
Assumption~\ref{below} hold.
Fix a reference probability $Q\in\tilde{\mathcal{P}}(W)$.
There exists $\phi\in\mathcal{A}$ such that
$V_T(z,\phi)\geq W$ a.s. if and only if
%
%e21 #&#
\begin{equation}
{Z}_0z\geq E_Q({Z}_TW)-E_Q\int
_0^T Z^0_t
G^*_t(\bar{Z}_t-S_t)\,dt,
\end{equation}
for all $\mathbb{R}^{d+1}_+$-valued bounded $Q$-martingales $Z$ with $Z_0^0=1$
satisfying $Z^i_t=0$, $i=1,\ldots,d$ on $\{Z^0_t=0\}$.
\end{theorem}

Defining $dQ'/dQ:=Z^0_T$ one can state Theorem~\ref{superreplication3}
in the following form, in which martingale probabilities $Q$ are
replaced by stochastic discount factors $Z$.

%co3.10 #&#
\begin{corollary}\label{superreplication4}
Let $W\in L^0(\mathbb{R}^{d+1})$, $z\in\mathbb{R}^{d+1}$ and
Assumption~\ref{below} hold.
Fix a reference probability $Q\in\tilde{\mathcal{P}}(W)$. There exists
$\phi\in\mathcal{A}$ such that
$V_T(z,\phi)\geq W$ a.s. if and only if
%
%e22 #&#
\begin{equation}
\hat{Z}_0z\geq E_{Q'}(\hat{Z}_T
W)-E_{Q'}\int_0^T
G^*_t({Z}_t-S_t)\,dt,
\end{equation}
for all $Q'\ll P$ with bounded $dQ'/dQ$ and for all $\mathbb
{R}^{d}_+$-valued $Q'$-martingales
$Z$ such that $(dQ'/dQ) Z_T$ is bounded.
\end{corollary}

Finally, in the case of a finite $\Omega$ Theorem~\ref
{superreplication3} reduces to a simple version, without any
integrability conditions.

%th3.11 #&#
\begin{theorem}\label{superreplication5} Let $\Omega$ be finite.
Let $W\in L^0(\mathbb{R}^{d+1})$, $z\in\mathbb{R}^{d+1}$ and let
Assumption~\ref{below} hold.
Fix any reference probability $Q\sim P$. There exists $\phi\in
\mathcal
{A}$ such that
$V_T(z,\phi)\geq W$ a.s. if and only if
%
%e23 #&#
\begin{equation}
{Z}_0z\geq E_Q({Z}_TW)-E_Q\int
_0^T Z^0_t
G^*_t(\bar{Z}_t-S_t)\,dt,
\end{equation}
for all $\mathbb{R}^{d+1}_+$-valued $Q$-martingales $Z$ with
$Z_0^0=1$, and
satisfying $Z^i_t=0$, $i=1,\ldots,d$ on $\{Z^0_t=0\}$.
\end{theorem}

\begin{pf*}{Proof of Theorem~\ref{superreplication2}}
For a $(d+1)$-dimensional vector $x$, $\tilde{x}$ denotes the
$d$-dimensional vector $\tilde{x}^i:=x^i$, $i=1,\ldots,d$ (cf. Table~\ref{tab:not}).
First, assume that $V_T(z,\phi)\geq W$. Take $Q\in\mathcal{P}(W)$
and a
bounded $Q$-martingale $Z$ with nonnegative components [more
generally, it is enough
to assume that $Z_T W$ is $Q$-integrable and that $Z_T\in L^{\gamma
}(Q)$], satisfying
$Z^i_t=0$, $i=1,\ldots,d$ on $\{Z^0_t=0\}$.

Note that $E_Q|W|<\infty$ and $W^0\leq z+\int_0^T  [-\phi_t
S_t-G_t(\phi_t) ]\,dt$ because\break $V_T(z,\phi)\ge W$, hence Lemma~\ref
{morecamb} implies
%
%e24 #&#
\begin{equation}
\label{kinross} E_Q\int_0^T |
\phi_t|^{\beta}\bigl(1+|S_t|\bigr)^{\beta}\,dt<\infty.
\end{equation}
Again, since $V_T(z,\phi)\ge W$, it follows that
%
%e25 #&#
\begin{equation}
{Z}_T(W-z)\leq\int_0^T
\bigl[-Z^0_T \phi_t S_t-Z^0_T
G_t(\phi_t) +\tilde{Z}_T\phi_t
\bigr] \,dt.
\end{equation}
By \eqref{kinross},
Fubini's theorem applies and the properties of conditional expectations
imply that
\begin{eqnarray*}
E_Q({Z}_TW)&\leq& z E_Q{Z}_T
+E_Q\int_0^T \bigl[-Z^0_T
\phi_t S_t-Z^0_T G_t(
\phi_t)+\tilde{Z}_T\phi_t \bigr] \,dt
\\
&=& z{Z}_0+\int_0^T
E_Q\bigl(-Z^0_T\phi_t
S_t-Z^0_T G_t(
\phi_t)+\tilde {Z}_T\phi_t\bigr)\,dt
\\
&=& z{Z}_0+ \int_0^T
E_Q\bigl(-Z^0_t\phi_t
S_t-Z^0_t G_t(
\phi_t)+\tilde {Z}_t\phi_t\bigr)\,dt
\\
&=& z{Z}_0+ \int_0^T
E_Q\bigl(-Z^0_t\phi_t
S_t-Z^0_t G_t(
\phi_t)+Z^0_t\bar {Z}_t
\phi_t\bigr)\,dt
\\
&\le& z{Z}_0+E_Q\int_0^T
Z^0_t G^*_t(\bar{Z}_t-S_t)
\,dt,
\end{eqnarray*}
which proves the first implication of this theorem.

To prove the reverse implication, suppose there is no $\phi$ such that
$V_T(z,\phi)\geq W$, which means that ${W}-z\notin C$. Fix $Q\in
\tilde
{\mathcal{P}}(W)$. The set $C\cap L^1(Q)$ is closed in $L^1(Q)$ by
Proposition~\ref{key}. The Hahn--Banach theorem then provides
a nonzero, bounded $(d+1)$-dimensional random variable $\tilde{Z}$
such that
%
%e26 #&#
\begin{equation}
\label{seti} E_Q\bigl[\tilde{Z}({W}-z)\bigr]>\sup
_{X\in C\cap L^1(Q)}E_Q[\tilde{Z} X].
\end{equation}

Since $-L^0(\mathbb{R}^{d+1})\subset C$,
$\tilde{Z}\geq0$ a.s., otherwise the supremum would be infinity. Define
now the (deterministic) processes $\psi(n,i)$ for all $n\in\mathbb{N}$
and $i=1,\ldots,d$ by setting
$\psi_t^i(n,i):=n$, $\psi_t^j(n,i)=0$, $j\neq i$ for all $t\in[0,T]$.

We claim that $E_Q\tilde{Z}^0>0$. Otherwise, for some $i>0$ one should
have $E_Q\tilde{Z}^i>0$. By Assumption~\ref{below} $\psi(n,i)\in
\mathcal{A}$.
By the choice of $Q$, we even have $V_T(0,\psi(n,i))\in C\cap L^1(Q)$
and $E_Q\tilde{Z}V_T(0,\psi(n,i))=nT E_Q\tilde{Z}^i\to\infty$ as
$n\to
\infty$, which is impossible
by \eqref{seti}. So we conclude that $E_Q\tilde{Z}^0>0$. Up to a
positive multiple of $Z$, we may assume
$E_Q\tilde{Z}^0=1$. Define $Z_t:=E_Q[\tilde{Z}|\mathcal{F}_t]$,
$t\in[0,T]$.

%Consider also $$
%Z^i_T:=\frac{\tilde{Z}^i}{\tilde{Z}^0}1_{\{\tilde{Z}^0\neq0\}},\ i=1,
%\ldots,d,
%$$
%and
%Define the martingales $\tilde{Z}_t:=E_{Q}[\tilde{Z}\vert
%\mathcal{F}_t]$, $t\in[0,T]$.
We also claim that, for all $i=1,\ldots,d$,
%
%e27 #&#
\begin{eqnarray}
\label{tarte} (P\times \operatorname{Leb}) (A_i)=0
\nonumber
\\[-8pt]
\\[-8pt]
\eqntext{\mbox{where }A_i:=
\bigl\{(\omega,t)\dvtx {Z}^0_t(\omega )=0\bigr\} \setminus
\bigl\{(\omega,t)\dvtx {Z}^i_t(\omega)=0\bigr\}.}
\end{eqnarray}
If this were not the case for some $i$, define $\psi^i(n,i):=n1_{A_i}$,
$\psi^j(n,i):=0$, $j\neq i$. Clearly, $\psi(n,i)\in\mathcal{A}$ and
$V_T(0,\psi(n,i))\in C\cap L^1(Q)$
while $E_{Q}\tilde{Z} V_T(0,\psi(n,i))\to\infty$, $n\to\infty$, which
is absurd, proving \eqref{tarte}.
%We define
%$$
%Z_t^i:=1_{\{\tilde{Z}_t^0\neq0\}}\frac{\tilde{Z}_t^i}{\tilde{Z}_t^0},
%\ i=1,\ldots,d.
%$$
%We know from \eqref{tarte} that $Z_t^i \tilde{Z}^0_t=\tilde{Z}^i_t$
%a.s. which is easily
%seen to imply that $Z_t$ is a $Q'$-martingale.

By the measurable selection theorem applied to the measure space
$(\Omega\times[0,T],\mathcal{O},P\otimes \operatorname{Leb})$ [see Proposition III.44
in \citet{dma}],
there is an optional process $\tilde{\chi}(n)$ such that
%
%e28 #&#
\[
 \tilde{\chi}_t(n)[\bar{Z}_t-S_t]-G_t
\bigl(\tilde{\chi}_t(n)\bigr)\leq G^*_t(
\bar{Z}_t-S_t)
\]
and
\begin{equation}
\label{malina}
\tilde{\chi}_t(n)[\bar{Z}_t-S_t]-G_t
\bigl(\tilde{\chi}_t(n)\bigr)\geq G^*_t(
\bar{Z}_t-S_t)-\frac{1}{n}\geq-K-\frac{1}{n},
\end{equation}
for $(P\times \operatorname{Leb})$-almost every $(\omega,t)$. Here $K$ denotes the
bound for\break $\sup_{t\in[0,T]} G_t(0)$ from \eqref{eq:0b}. Now define
$\chi
_t(n):=\tilde{\chi}_t(n)
1_{\{|\tilde{\chi}_t(n)|\leq N(n)\}}$ where $N(n)$ is chosen such that
$(P\times \operatorname{Leb})(|\tilde{\chi}_t(n)|>N(n))\leq1/n^2$.
By Assumption~\ref{below}, $\chi(n)\in\mathcal{A}$ and
by the choice of $Q$, $V_T(0,\chi(n))\in C\cap L^1(Q)$. By construction,
\[
\lim_{n\rightarrow\infty} \chi_t(n)[\bar{Z}_t-S_t]-G_t
\bigl(\chi_t(n)\bigr)= G^*_t(\bar{Z}_t-S_t),\qquad
(P\times \operatorname{Leb})\mbox{-a.e.}
\]
%
%Note that if
%$$
%|x|>\left(\frac{|y|}{H_t}\right)^{1/(\alpha-1)}
%$$
%then $yx-G_t(x)\leq yx-H_t|x|^{\alpha}<0$. This implies that
%$$
%|\phi_t(n)|\leq\left(\frac{|{Z}_t-S_t|}{H_t}\right)^{1/(\alpha-1)}
%$$
%almost surely. Here the numerator is a c\`adl\'ag process and the
%denominator is
%bounded away from $0$ a.s. by Assumption~\ref{below}. It follows that
%the function $|\phi_t(n)|$,
%$t\in[0,T]$ is a.s. bounded which implies $\int_0^T |\phi_t(n)|\,dt<
%\infty$ a.s., hence
%$\phi(n)\in\mathcal{A}$.

%Since $Z_T$ is bounded, the lower bound in \eqref{malina} allows the
%use of Fubini's theorem and
%Fatou's lemma, hence
Since $Z$, $\chi(n)$ are bounded and $Q \in \tilde{\mathcal{ P}}$
we may use Fubini's theorem and
the lower bound in (\ref{malina}) allows the use of Fatou's lemma, hence
\begin{eqnarray*}
\liminf_{n\to\infty} E_QZ_T V_T
\bigl(0,\chi(n)\bigr) &=&\liminf_{n\to\infty}E_Q\int
_0^T \chi_t(n)\bigl[
\tilde{Z}_T-Z^0_TS_t
\bigr]-Z^0_TG_t\bigl(\chi_t(n)
\bigr)\,dt
\\
&=&
\liminf_{n\to\infty}\int
_0^TE_{Q} \bigl[{\chi}_t(n)\bigl[
\tilde{Z}_t-Z_t^0 S_t
\bigr]-Z_t^0 G_t\bigl({\chi}_t(n)
\bigr)\bigr]\,dt
\\
&=&
\liminf_{n\to\infty}E_{Q}\int
_0^T{\chi }_t(n)Z_t^0[
\bar {Z}_t- S_t]-Z_t^0
G_t\bigl({\chi}_t(n)\bigr)\,dt
\\
&\ge& E_{Q}\int_0^T
Z_t^0 G^*_t(\bar{Z}_t-S_t)
\,dt.
%
%\label{torrero}
\end{eqnarray*}
%
%as $n\to\infty$.

From \eqref{seti}, we infer that
\begin{eqnarray*}
z{Z}_0&<& \limsup_{n\to\infty} \bigl[E_{Q}(W{Z}_T)-E_{Q}{Z}_T
V_T\bigl(0,\chi (n)\bigr) \bigr]\\
&=&
E_{Q}(W{Z}_T)-\liminf_{n\to\infty}
E_{Q}{Z}_T V_T\bigl(0,\chi(n)\bigr)\\
&\leq&
E_{Q}(W{Z}_T)-E_{Q}\int_0^T
Z^0_t G^*_t(\bar{Z}_t-S_t)
\,dt.
\end{eqnarray*}
This completes the proof.
\end{pf*}

%re3.12 #&#
\begin{remark}\label{torun}
The above proof also shows that the statements of Theorems~\ref
{superreplication2} and \ref{superreplication3} remain valid when the
class of bounded martingales is replaced by
the class of $Q$-martingales with $Z_T\in L^{\gamma}(Q)$ such that
${Z}_T W$ is $Q$-integrable.
\end{remark}

For a real number $c$, denote by $\check{c}$ the $(d+1)$-dimensional
vector $(c,0,\ldots,0)^{T}$ (cf. Table~\ref{tab:not}). The next
corollary specializes Theorem~\ref{superreplication2} to the situation
in which a claim in cash is hedged from an initial cash position only.

%co3.13 #&#
\begin{corollary}\label{superreplication6}
Let $W\in L^0(\mathbb{R})$, $c\in\mathbb{R}$ and let Assumption~\ref
{below} hold.
There exists $\phi\in\mathcal{A}$ such that
$V_T^0(\check{c},\phi)\geq W$ a.s. and $V_T^i(\check{c},\phi)\geq0$,
$i=1,\ldots,d$ if and only if
%
%e29 #&#
\begin{equation}
\label{mabel} c\geq E_Q\bigl(Z^0_T W
\bigr)-E_Q\int_0^T
Z^0_t G^*_t(\bar{Z}_t-S_t)
\,dt,
\end{equation}
for all $Q\in\mathcal{P}(W)$ and for all
$\mathbb{R}^{d+1}_+$-valued bounded $Q$-martingales $Z$ with $Z^0_0=1$
satisfying $Z^i_t=0$, $i=1,\ldots,d$ on $\{Z^0_t=0\}$.
\end{corollary}

To understand the meaning of \eqref{mabel}, it is helpful to consider
its statement in the frictionless case, at least formally.\footnote{The theorem does not apply to the frictionless case because $G=0$ does not
satisfy Assumption~\ref{below}, and feasible strategies differ from
admissible strategies.} If $S$ itself is a $Q$-martingale, then the
penalty term with $G^*$ vanishes with the choice of $Z_t^0:=1$,
$Z_t^i:=S_t^i$, $i=1,\ldots,d$. It follows that, in order to
super-replicate $W$, the initial endowment $c$ must be greater than or
equal to the supremum of $E_Q W$ over the set of equivalent martingale
measures for $S$. This shows that our findings are formally consistent
with well-known superhedging theorems for frictionless markets. The
results are similarly consistent with superhedging theorems for
proportional transaction costs [\citet{MR2589621}], formally obtained
with $G_t(x) = \varepsilon S_t |x|$.

%s3.5 #&#
\subsection{Examples}

With the class of superlinear frictions considered in this article,
typical contingent claims
are virtually impossible to superreplicate with certainty at a fixed
price, as we now show.
For example, consider the problem of delivering a cash payoff equal to
$S_T$ (the price of
the risky asset) at time $T$, starting from cash only. In a market
without frictions, or with
proportional transaction costs, one solution is to immediately buy the share
and, therefore, the superreplication price is at most the (current)
price of the asset (or a slightly higher
multiple when transaction costs are present).

But this policy is not feasible with superlinear frictions, as block
trades are forbidden. An approximate solution would be to buy at rate
$n$ over the period $[0,1/n]$, but this policy incurs a positive
probability that the asset price will rapidly increase in value, and in
typical models, such as geometric Brownian motion, there is no certain
upper bound on the execution price.

This discussion motivates the following result.

%ex3.14 #&#
\begin{example} Let $\mu\in\mathbb{R}$, $\sigma,S_0>0$,
$S_t:= S_0 e^{(\mu-\sigma^2/2)t+\sigma W_t}$, $G_t(x) = \frac
{\lambda
}{2}S_t x^2$,
where $W_t$ is a Brownian motion (and $\mathcal{F}_t$ is its completed
filtration made
right-continuous). Then a cash payoff equal to $S_T$ cannot be
superreplicated from any initial capital.
\end{example}

\begin{pf}
In view of Theorem~\ref{superreplication2} above, it is enough to show
that the right-hand side of
inequality \eqref{viola} takes arbitrarily large values for a suitable
family of reference probabilities $Q$ and martingales $Z$.

To this end, consider $Q = P$ and the family of exponential martingales
$Z$ parameterized by $x>0$
and $n\in\mathbb{N}$, $n>1/T$ with
%
%e30 #&#
\begin{eqnarray}
&&Z^0_t = \exp\biggl\{-\sigma W_{t\wedge(T-1/n)} - \frac{\sigma^2}2 t\wedge
(T-1/n)
\nonumber
\\[-8pt]
\\[-8pt]
\nonumber
&&\hspace*{46pt}{}+
1_{\{t\ge T-1/n\}}  \biggl( (x-\sigma) (W_t-W_{T-1/n})-\frac{(x-\sigma
)^2}2 \bigl(t-(T-1/n)\bigr)  \biggr)\biggr\}\hspace*{-30pt}
\end{eqnarray}
and $Z^1_t = S_0 Z^0_t$. [In plain English, $Z^0_t$ adds a drift of
$-\sigma$ (to the Brownian motion)
between $0$ and $T-1/n$, and a drift of $x-\sigma$ between $T-1/n$ and
$T$.] In the sequel,
$C_1,C_2,\ldots$ denotes various positive constants whose values do
not depend
either on $x$ or on $n$.

Notice that, for $0\leq t\leq T-1/n$,
\[
EZ^0_t S_t=S_0 e^{(\mu-\sigma^2)t}
\leq C_1
\]
and for $T-1/n\leq t\leq T$,
\begin{eqnarray*}
EZ^0_t S_t&\leq& C_1
e^{(x^2/2)(t-(T-1/n))+(\mu-\sigma
^2/2)(t-(T-1/n))-((x-\sigma)^2/2)(t-(T-1/n))}\\
&\leq& C_2 e^{\sigma x/n}.
\end{eqnarray*}

Similarly, for $0\leq t\leq T-1/n$,
\[
ES_0^2 Z^0_t/S_t=E S_0
e^{-2\sigma W_t-(\mu-\sigma^2/2)t-(\sigma
^2/2)t}\leq C_3
\]
and for $T-1/n\leq t\leq T$,
\begin{eqnarray*}
ES_0^2 Z^0_t/S_t&\leq&
C_3 e^{((x-2\sigma)^2/2)(t-(T-1/n))-(\mu-\sigma
^2/2)(t-(T-1/n))-
((x-\sigma)^2/2)(t-(T-1/n))}\\
&\leq& C_4.
\end{eqnarray*}

We also have
\[
EZ^0_T S_T\geq S_0
e^{(\mu-\sigma^2)(T-1/n)}e^{(x^2/2)(1/n)+(\mu
-\sigma
^2/2)(1/n)-((x-\sigma)^2/2)(1/n)}\geq C_5 e^{\sigma x/n}.
\]

Now set $x=x(n)=n\ln n/\sigma$.
Since $G_t^*(y) = \frac{1}{2\lambda S_t}y^2$, for $W = (S_T,0)$, which
represents a cash payoff equal to the final stock price, it follows that
%
%e31 #&#
\begin{eqnarray}
&&E({Z}_TW)-E\int_0^T
Z^0_t G^*_t(\bar{Z}_t-S_t)
\,dt\nonumber\\
&&\qquad = E\bigl[S_T Z^0_T\bigr]-
\frac
{1}{2\lambda} \int_0^TE \biggl[
\frac{(Z^1_t)^2}{S_t Z^0_t} -2 Z^1_t + Z^0_t
S_t \biggr]\,dt
\nonumber
\\
&&\qquad=  E\bigl[S_T Z^0_T\bigr] -
\frac{1}{2\lambda} \int_0^T \biggl( E \biggl[
\frac
{S_0^2 Z^0_t }{S_t} \biggr] -2 S_0 + E \bigl[Z^0_t
S_t \bigr] \biggr)\,dt\\
&&\qquad\geq C_5 n- \frac{1}{2\lambda} \int_0^{T-1/n}
[C_3-2S_0+C_1] \,dt-\frac
{1}{2\lambda n}
[C_4 -2S_0+C_2 n]\nonumber\\
&&\qquad\geq C_5 n-C_6\to\infty\nonumber %\ge& S_0 e^{\mu T+ \sigma x (T-1/n) + \log n)} \\ - & \frac{1}{2
%\lambda}
%\left(S_0 \int_0^{T-1/n} (e^{(-\mu+\sigma^2-\sigma x)t}-2 + e^{(\mu+
%\sigma x)t})\,dt+\frac1n
%\left( E\left[ \frac{S_0^2 Z^0_{T-1/n} }{S_{T-1/n}}\right] -2 S_0 + E
%\left[Z^0_T S_T\right] \right)
%( e^{(-\mu+\sigma^2+\sigma x(n+1))t} -2 + e^{(\mu+\sigma x (n+1))t} )\,dt
%\right)
\end{eqnarray}
%
%where the inequality holds because because $t\mapsto E \left[Z^0_t S_t
%\right]$ increases in $t$, while $t\mapsto E\left[ \frac{S_0^2 Z^0_t
%}{S_t}\right]$ decreases in $t$, since $x>0$.
%Now, note that the first term $E[S_T Z^0_T] = S_0 e^{\mu T+ \sigma x
%(T-1/n) + \log n}$ becomes arbitrarily high for $n$ large, while $
%\frac1n E[S_T Z^0_T] = S_0 e^{\mu T+ \sigma x (T-1/n)}$ remains
%bounded. Likewise, also $\frac1n E[S_0^2 Z^0_{T-1/n}/S_{T-1/n}] = S_0
%e^{(-\mu+ \sigma^2) {T-1/n} - \sigma x (T-1/n) -\log n}$ remains
%bounded.
as $n\to\infty$. As a result, the right-hand side takes arbitrarily
large values, implying an infinite superreplication price.
\end{pf}

The previous proof uses Theorem~\ref{superreplication2} to obtain a
dual characterization of superreplication prices. In fact, the same
conclusion can be reached exploiting the market bound obtained in Lemma~\ref{finite}.

\begin{pf*}{Alternative proof}
Observe that $G_t(x) = \frac\lambda2 S_t x^2$ implies that $G^*_t(y) =
\frac{y^2}{2\lambda S_t}$, whence the market bound is
%
%e32 #&#
\begin{equation}
B=\int_0^T G^*_t(-S_t)
\,dt = \frac{1}{2\lambda}\int_0^T
S_t \,dt.
\end{equation}
Thus, any strategy starting with initial capital $x$ satisfies the bound
%
%e33 #&#
\begin{equation}
V^0_T(x,\phi)\leq x+\int_0^T
G^*_t(-S_t)\,dt \le x + \frac{1}{2\lambda
}\int
_0^T S_t \,dt.
\end{equation}
In particular, on the event $ \{ x + \frac{1}{2\lambda}\int_0^T S_t
\,dt < S_T \}$, which has positive probability for any $x$ (because
Brownian motion has full support on the space of continuous functions
starting at $0$) superreplication fails for any strategy, and for any
initial capital.
\end{pf*}

The previous example should be understood as follows: if a large
position in the risky asset needs to be acquired, it is not possible a
priori to guarantee a fixed execution price with certainty: price
impact prevents the transaction to take place instantly, while over
time intervening news may lead the price to arbitrary levels. Yet, the
fact that even such a simple contract is not superreplicable with
finite wealth raises the question of which contracts have a finite
superhedging price, and the next example provides one.

%ex3.15 #&#
\begin{example}
Let $S_t >0$ a.s. for all $t$ and $G_t(x):= \frac\lambda2 S_t x^2$.
Then, for all $k>0$, the contract that at time $T$ pays $\frac{1}\lambda
\int_0^T (\sqrt{1+2 k \lambda/S_t}-1)\,dt$ units of the risky asset is
superreplicable from initial cash position $kT$.
\end{example}

\begin{pf}
The main idea is that this payoff is dominated by a \emph{constant
cash-flow} strategy, a strategy that buys the risky asset at the rate
of one unit of the safe asset per unit of time (e.g., one dollar per
second). To see this, recall the relation between the cash flow and the
trading rate
%
%e34 #&#
\begin{equation}
dV^0_t = -\phi_t S_t \,dt -\frac
\lambda2 S_t \phi_t^2 \,dt.
\end{equation}
Thus, a constant cash flow $dV^0_t = -k \,dt$ corresponds to a buying rate
%
%e35 #&#
\begin{equation}
\phi_t = \frac{1}\lambda \biggl(-1 + \sqrt{1+
\frac{2 \lambda
k}{S_t}} \biggr),
\end{equation}
which yields at time $T$ exactly
$\frac{1}\lambda\int_0^T  (-1 + \sqrt{1+\frac{2 \lambda
k}{S_t}} )\,dt$
units of the risky asset.
In the frictionless limit ($\lambda\downarrow0$), this strategy
implies a buying rate of $\phi_t = k/S_t$, which yields $k \int_0^T
1/S_t \,dt$ units of the risky asset.
\end{pf}

In the above example note that, as $k$ varies, the resulting family of
payoffs is not linear, in that while each of the above payoffs are
replicable, their multiples need not be. In particular, increasing the
buying rate $k$ does not scale the number of units of risky asset
bought proportionally, except in the frictionless limit $\lambda= 0$.
Note also that the above payoff is superreplicable because it promises
a lower number of shares when the asset price is high. The square-root
relation is of course linked to the quadratic price impact considered
in this example.

%s4 #&#
\section{Arbitrage (of the second kind)}\label{444}

Any positive payoff that is superhedged for strictly less than zero is
an arbitrage.
Such opportunities, which start from an insolvent position and, by
clever trading, yield a solvent one, are known in the literature as
arbitrage of
the second kind, and date back to \citet{i} [see also \citet{MR1348197}
in the context of large financial markets].
This definition is used with markets frictions in \citeauthor{dr1}
(\citeyear{dr1,dr2}), and,
more recently, in \citet{MR2648605,kd,bruno1,bruno2,teemu}.

The superhedging results in the previous section hold regardless of
having arbitrage opportunities or not.
Consequently, they can be used to \emph{detect} arbitrage: if we find a
nonnegative payoff $W$ satisfying \eqref{mabel} with some $c<0$ then
Corollary~\ref{superreplication6} ensures that an arbitrage
opportunity exists.

%de4.1 #&#
\begin{definition} An arbitrage of the second kind is a strategy $\phi
\in\mathcal{A}$, such that $V_T(\check{c},\phi)\geq0$ for some $c<0$.
Absence of arbitrage of the second kind (NA2) holds if no such
opportunity exists.
\end{definition}

Note that this definition requires that $S$ has positive components.
Otherwise, a~nonnegative position in an asset with negative price [as $V_T(\check
{c},\phi)\geq0$ stipulates] cannot be interpreted as solvent.

The following theorem is a direct consequence of Corollary~\ref
{superreplication6} and Remark~\ref{torun}.

%th4.2 #&#
\begin{theorem}\label{char}
Let Assumption~\ref{below} hold. Then (NA2) holds if and only if,
for all $\varepsilon>0$, there exists $Q\in\mathcal{P}$ and an
$\mathbb
{R}^{d+1}_+$-valued
$Q$-martingale $Z$ with $Z_T\in L^{\gamma}(Q)$ such that $E_Q\int_0^T
Z^0_t G_t^*(\bar{Z}_t-S_t)\,dt<\varepsilon$.
\end{theorem}

A broad class of models enjoys the (NA2) property.
Let $D\subset(0,\infty)^d$ be nonempty, open and convex. We denote by
$C[t,T](D)$ (resp., $C_x[t,T](D)$) the set of continuous functions $f$ from
$[t,T]$ to $D$ [resp., satisfying $f(t)=x$]. Both spaces are equipped
with the Borel sets of the topology induced by the uniform metric.
Recall that a continuous stochastic process $S$ on $[t,T]$ can be
understood as a $C[t,T](D)$-valued random variable, and its support is
defined in this (metric) space.

%de4.3 #&#
\begin{definition}
A process $S$ has conditional full support in $D$ (henceforth, CFS-$D$) if
$S\in C[0,T](D)$ a.s. and
\[
\operatorname{supp}P(S|_{[t,T]}\in\cdot\vert\mathcal{F}_t)=C_{S_t}[t,T](D)\qquad
\mbox{a.s. for all }t\in[0,T].
\]
\end{definition}

%th4.4 #&#
\begin{theorem}\label{suff} Let Assumption~\ref{below} hold with
$H_t:=H$ constant.
If $S$ has the \textup{CFS}-$D$ property, then {\textup{(NA2)}} holds.
\end{theorem}

\begin{pf}
%Let us fix $Q'\sim P$ such that $L_N\in L^2(Q')$ for all $N$
%(such a $Q'$ exists by p. 266 of \citet{dm2}). CFS-$D$ continues to
%hold under $Q'$.
It follows from Theorem~2.6 of \citet{sayit} that for all $\varepsilon$
there is $Q\sim P$ and a
$Q$-martingale $M_t$ evolving in $D\subset\mathbb{R}^d_+$ such that
$|S_t-M_t|<\varepsilon$ a.s. for all $t$. Define $Z_t^i:=M_t^i$
for $i=1,\ldots,d$ and $Z^0_t:=1$ for all $t$.

%It follows that $|S_t-\tilde{Z}_t|\leq2\varepsilon$ for the $t\in
%[0,T]$.
%By convexity of $D$, $\tilde{Z}_t\in D\subset\mathbb{R}_+^d$

In \citet{sayit} [see also \citet{grs}] it is shown that $S_T$, and hence
${Z}_T$ are in $L^2(Q)$. A closer inspection of the proof reveals that
in fact there exist ${Z}_T\in L^p(Q)$ for arbitrarily large $p$. Take
$p:=\max\{\gamma, \alpha\beta/(\alpha-\beta)\}$.
Then $Q$ is easily seen to be in $\mathcal{P}$ and ${Z}_T$ is in
$L^{\gamma}(Q)$.
%It follows that $|E_Q(\tilde{S}_t-S_t)|\leq2\varepsilon$ a.s.
%Choose $Z_t\equiv0$.
The estimate \eqref{eq:star} in Lemma~\ref{b} implies that
\[
E_Q\int_0^T G_t^*(
\bar{Z}_t-S_t)\,dt=E_Q\int
_0^T G_t^*({M}_t-S_t)
\,dt\leq \int_0^T \ell(\varepsilon)\,dt\leq T
\ell(\varepsilon)
\]
for a continuous (deterministic) function $\ell$, which clearly tends
to $0$ as $\varepsilon\to0$. Now the claim follows by Theorem~\ref{char}.
\end{pf}

Theorem~\ref{suff} has an immediate implication for fractional
Brownian motion.
The arbitrage properties of fractional Brownian motion have long been
delicate: in a frictionless setting it admits arbitrage of the second
kind [\citet{rogers1997arbitrage}] but, with proportional transaction
costs, it does not even have arbitrage of the first kind [\citet{grs}].
With price-impact, the above theorem implies that it does not admit
arbitrage of the second kind, since it satisfies the CFS-$D$ property
[\citet{grs}].
Whether arbitrage of the first kind (a positive, and possibly strictly
positive, payoff from nothing) exists is still an open question.

%s5 #&#
\section{Utility maximization}\label{555}

This section discusses utility maximization in the model of Section~\ref{222}. The first result (Theorem~\ref{utility} below) shows that
optimal strategies exist under a simple integrability assumption, which
is easy to check in practice. In particular, optimal strategies exist
regardless of arbitrage, since such opportunities are necessarily
limited. Put differently, the budget equation is nonlinear. Therefore,
one cannot add to an optimal strategy an arbitrage opportunity, and
expect the resulting wealth to be the sum.

The second result establishes the first-order condition for utility
maximization, which provides a simple
criterion for optimality, and helps understand the differences with the
corresponding results for
frictionless markets and proportional transaction costs. In particular,
it shows that the analogue of a shadow price for price-impact models is
a hypothetical frictionless price for which the optimal strategy would
coincide with the execution price of the same strategy in the original
price-impact model.
This notion reduces to that of shadow price for markets with proportional
transaction costs.

Importantly, these results consider only utilities defined on the real
line, such as exponential utility, but
exclude power and logarithmic utilities, which are defined only for
positive values. This setting is consistent with the definition of
feasible strategies, which do not constrain wealth to remain positive.
When establishing optimality of a given strategy
in such a setting, one technical challenge is to show that the
resulting wealth processes are martingales (or just supermartingales)
with respect to appropriate reference measures
(these are martingale measures in the frictionless case). Lemma~\ref
{martingale} below implies such a property
for any feasible strategy and hence forms the main ingredient of the
proof of Theorem~\ref{bagdad}.
Finally, since the focus is on utility functions defined on a single
variable, and with price impact there is no scalar notion of portfolio
value, the results below assume for
simplicity that all strategies begin and end with cash only.

Let $W$ be an arbitrary real-valued random variable (representing a
random endowment) and $c\in\mathbb{R}$ the investor's initial capital.

%th5.1 #&#
\begin{theorem}\label{utility} Let $U\dvtx \mathbb{R}\to\mathbb{R}$ be
concave and nondecreasing,
and let $E|U(c+B+W)|<\infty$ hold
for the market bound $B=\int_0^T G^*_t(-S_t)\,dt$ in Lemma~\ref{b}. Under
Assumption~\ref{below}, there is $\phi^*\in\mathcal{A}'(U,c)$
such that
\[
EU\bigl(V_T^0\bigl(\check{c},\phi^*\bigr)+W\bigr)=\sup
_{\phi\in\mathcal
{A}'(u,c)}EU\bigl(V_T^0(\check{c},\phi)+W
\bigr),
\]
where $
\mathcal{A}'(U,c)=\{\phi\in\mathcal{A}\dvtx V_T^i(\check{c},\phi)=0,
i=1,\ldots,d,
EU_-(V_T^0(\check{c},\phi)+W)<\infty\}$.
\end{theorem}

This theorem applies, in particular, for $U$ bounded above and $W$
bounded below.

\begin{pf*}{Proof of Theorem \ref{utility}} Corollary~\ref{closed} implies that
\[
C':=\check{c}+ \bigl(C\cap\bigl\{X\dvtx X^i=0\mbox{
a.s.}, i=1,\ldots,d\bigr\} \bigr)
\]
is closed in probability.

Let $\phi(n)$ be a sequence in $\mathcal{A}'(U,c)$ with
\[
\lim_{n\to\infty}EU\bigl(V_T^0\bigl(
\check{c},\phi(n)\bigr)+W\bigr)=\sup_{\phi\in
\mathcal
{A}'(U,c)}EU
\bigl(V_T^0(\check{c},\phi)+W\bigr).
\]
Since $V_T^0(\check{c},\phi(n))\leq c+B$ a.s. for all $n$, by Lemma~9.8.1 of \citet{dsc} there are convex combinations
such that
$\sum_{j=n}^{M(n)} \alpha_j(n)\* V_T^0(\check{c},{\phi}(j))\to V$ a.s.
for some $[-\infty,c+B]$-valued random variable
$V$. By convexity of $G$, we have that
for $\tilde{\phi}(n):=\sum_{j=n}^{M(n)}\alpha_j(n) \phi(j)$,
\[
V_T^0\bigl(\check{c},\tilde{\phi}(n)\bigr)\geq\sum
_{j=n}^{M(n)} \alpha_j(n)
V_T^0\bigl(\check{c},{\phi}(j)\bigr),
\]
so $\sum_{j=n}^{M(n)} \alpha_j(n) V_T(\check{c},\phi(j))\in C'$ for
each $n$.

By the concavity of $U$,
\[
EU \Biggl(W+\sum_{j=n}^{M(n)}
\alpha_j(n) V_T^0\bigl(\check{c},{\phi }(j)
\bigr) \Biggr)\geq\sum_{j=n}^{M(n)}
\alpha_j(n) EU\bigl(V_T^0\bigl(\check{c},{
\phi}(j)\bigr)+W\bigr).
\]

Fatou's lemma implies that $EU(V)\geq\sup_{\phi\in\mathcal
{A}'(u)}EU(V_T^0(\check{c},\phi)+W)$,
in particular, $V$ is finite-valued and hence $\check{V}\in C'$ by the
convexity and closedness of~$C'$.
It follows that $V=V_T^0(\check{c},\phi^*)-Y^0$ for some
$\phi^*\in\mathcal{A}'(U,c)$ and $Y\in L_+^0$. Clearly,
$EU(V_T^0(\check{c},\phi^*)+W-Y^0)=\sup_{\phi\in\mathcal
{A}'(U,c)}Eu(V_T^0(\check{c},\phi)+W)$.
Necessarily, $EU(V_T^0(\check{c},\phi^*)+W)=\sup_{\phi\in\mathcal
{A}'(U,c)}EU(V_T^0(\check{c},\phi)+W)$
as well.\footnote{Note that $U$ can be constant on an (infinite)
interval hence $Y^0\neq0$
is possible.} This completes the proof.
\end{pf*}

%re5.2 #&#
\begin{remark}
Theorem~\ref{utility} can also be proved with
\[
\mathcal{A}''(U,c)=\bigl\{\phi\in\mathcal{A}\dvtx
V_T^i(\check{c},\phi)\geq 0, i=1,\ldots,d, EU_-
\bigl(V_T^0(\check{c},\phi)+W\bigr)<\infty\bigr\}
\]
in lieu of $\mathcal{A}'(U,c)$. Note that the two optimization problems
are \emph{not} equivalent, due to illiquidity.
\end{remark}

%re5.3 #&#
\begin{remark}\label{remi2} Let us assume that $S$ is nonnegative
and one-dimensional. We may
replace $\mathcal{A}'(U,c)$ in Theorem~\ref{utility} by
\begin{eqnarray*}
\mathcal{A}'_+(U,c)&:=& \bigl\{ \phi\in\mathcal{A}\dvtx S_t(\omega) +
G\bigl(\omega,t,\phi_t(\omega)\bigr)/\phi_t(\omega)\geq0
\mbox{ when }\phi_t(\omega )\neq0,
\\
&&\hspace*{28pt} V_T^i(\check{c},\phi)\geq0, i=1,\ldots,d, EU_-
\bigl(V_T^0(\check{c},\phi)+W\bigr)<\infty\bigr\},
\end{eqnarray*}
that is, we may restrict our class of strategies to those for which the
instantaneous execution price
is nonnegative, as in Remark~\ref{remi} above.
\end{remark}

%re5.4 #&#
\begin{remark}
The proofs of Theorem~\ref{utility} and Proposition~\ref{key} use
Lemmata~9.8.1 and 15.1.4 in \citet{dsc}. They could be replaced, with
minor modifications, with Koml\'os's theorem [\citet{komlos}] and its
extensions [\citet{balder,vw}].
\end{remark}

While the previous result shows the existence of optimal strategies,
the next theorem provides a sufficient conditions for a strategy's
optimality, through a variant of the usual first-order condition.

%th5.5 #&#
\begin{theorem}\label{bagdad} Let Assumption~\ref{below} hold, and
\begin{enumerate}[(a)]
\item[(a)]
let $U$ be concave, continuously differentiable, with $U'$ strictly
decreasing, and
%
%e36 #&#
\begin{equation}
\label{luk} U(x)\leq-C|x|^{\delta}, \qquad x\leq0,
\end{equation}
for some $C>0$ and $\delta>1$;
\item[(b)]
denoting by $\tilde{U}$ the convex conjugate function of $U$, that is,
\[
\tilde{U}(y):=\sup_{x\in\mathbb R} \bigl\{U(x)-xy\bigr\},\qquad y>0;
\]
%
%assume that $\tilde{U}'$ exists and it is strictly increasing;
%
\item[(c)]
let $W$ be a bounded random variable;
\item[(d)]
let $Q\in\mathcal{P}$ be such that
%
%e37 #&#
\begin{equation}
\label{lak} dQ/dP\in L^{\eta}(P),
\end{equation}
where $(1/\eta) +(1/\delta)=1$;
\item[(e)]
let $G_t(\cdot)$ be $P\times \operatorname{Leb}$-a.s. continuously differentiable in
$x$ and $G_t'(\cdot)$ is strictly increasing;
\item[(f)]
let $Z$ be a c\`adl\`ag process with $Z_T\in L^{\gamma'}(Q)$ for some
$\gamma'>\gamma$ and
let ${\phi}^*$ be a feasible strategy such that, for some $y^*>0$, the
following conditions hold:
\begin{enumerate}[(i)]
\item[(i)]$Z$ is a $Q$-martingale;

\item[(ii)]$U'(V_T^0(x,{\phi}^*)+W) = y^* (dQ/dP)$ a.s.;

\item[(iii)]$Z_t = S_t + G'_t(\phi^*_t)$ a.s. in $P\times
 \operatorname{Leb}$-a.e.% for some
%admissible strategy $\hat\phi$;
%\item[(iv)]$E_Q (V_T^0(x,\phi^*) - \int_0^T G_t^*(Z_t - S_t) \,dt
% )= x$.
\end{enumerate}
\end{enumerate}

Then the strategy $\phi^*$ is optimal for the problem
%
%e38 #&#
\begin{equation}
\max_{\phi\in\mathcal{A}'(U,c)}E\bigl[U\bigl(V_T^0(x,
\phi)+W\bigr)\bigr].
\end{equation}
\end{theorem}

\begin{pf}
For any $(\phi_t)_{t\ge0}\in\mathcal{A}'(U,c)$ the final payoff equals
%
%e39 #&#
\begin{equation}
V_T^0(x,\phi) = x - \int_0^T
S_t \phi_t \,dt -\int_0^T
G_t(\phi_t) \,dt.
\end{equation}
Let $Z_t$ be as in the statement of the theorem, and rewrite the above
payoff as
\[
V_T^0(x,\phi) = x - \int_0^T
Z_t \phi_t \,dt + \int_0^T
(Z_t - S_t) \phi _t \,dt -\int
_0^T G_t(\phi_t) \,dt.
%\\
%=& x + \int_0^T Z_t\phi_t \,dt + \int_0^T (Z_t - S_t) \phi_t \,dt -
%\int_0^T G_t(\phi_t) \,dt.
\]
By definition of $G_t^*$, it follows that
%
%e40 #&#
\begin{equation}
\label{qwertz} V_T^0(x,\phi) \le x - \int
_0^T Z_t\phi_t \,dt +
\int_0^T G_t^*(Z_t -
S_t) \,dt,
\end{equation}
and equality holds if $Z_t - S_t = G'_t(\phi_t)$, $P\times \operatorname{Leb}$-a.s.,
that is, when (iii) holds.

It follows from Lemma~\ref{martingale} that
%
%e41 #&#
\begin{equation}
\label{eq:superr} 0 \le E_Q \biggl[ \biggl(x- V_T^0(x,
\phi) +\int_0^T G_t^*(Z_t
- S_t) \,dt \biggr) \biggr].
\end{equation}
Thus, for any payoff $V_T^0(x,\phi)+W$ and any $y>0$ the following holds:
%
%e42 #&#
\begin{eqnarray}\label{mobel}
&&E\bigl[U\bigl(V_T^0(x,\phi)+W\bigr)\bigr]\nonumber\\
&&\qquad\le E
\biggl[U\bigl(V_T^0(x,\phi)+W\bigr)
\nonumber
\\[-8pt]
\\[-8pt]
\nonumber
&&\hspace*{45pt}{} + y (dQ/dP)
\biggl(x-V_T^0(x,\phi) +\int_0^T
G_t^*(Z_t - S_t) \,dt \biggr)\biggr]
\\
&&\qquad\le E\biggl[\tilde{U}\bigl(y (dQ/dP)\bigr)+y (dQ/dP) \biggl(\int
_0^T G_t^*(Z_t -
S_t) \,dt + W \biggr) \biggr]+ y x.\nonumber
\end{eqnarray}
If (iii) is satisfied then there is equality in \eqref{qwertz} above.
If, in addition, (ii) is satisfied then both inequalities
in \eqref{mobel} are equalities for $y=y^{*}$.
%Since \eqref{mobel}
%holds for any $y>0$, it follows that
%%
%%e43 #&#
%\begin{eqnarray}
%\label{qkac} &&\sup_{\phi\in\mathcal{A}'(U,c)} E\bigl[U\bigl(V_T^0(x,
%\phi)+W\bigr)\bigr]\nonumber\\
%&&\qquad\le \inf_{y>0} \biggl(E\biggl[\tilde{U}
%\bigl(y (dQ/dP)\bigr)\\
%&&\hspace*{68pt}{}+y (dQ/dP) \biggl(\int_0^T
%G_t^*(Z_t - S_t) \,dt + W \biggr)\biggr]+ y x
%\biggr).\nonumber
%\end{eqnarray}
%%
%The infimum on the right-hand side is achieved at $y^*$ if the
%following condition holds:
%%
%%e44 #&#
%\begin{equation}
%E_Q \biggl[-\tilde{U}'\bigl(y^* (dQ/dP)\bigr)-
%\biggl(\int_0^T G_t^*(Z_t
%- S_t) \,dt + W \biggr) \biggr] = x.
%\end{equation}
%%
%Since $-\tilde{U}' = (U')^{-1}$, the above condition, combined with (ii),
%reduces to
%%
%%e45 #&#
%\begin{equation}
%E_Q \biggl(V_T^0(x,\phi) - \int
%_0^T G_t^*(Z_t -
%S_t) \,dt \biggr) = x,
%\end{equation}
%%
%which coincides with condition (iv).
Thus, if conditions (i), (ii) and
(iii) hold for ${\phi}^*$ then, by \eqref{mobel},
\begin{eqnarray*}
&&E\bigl[U\bigl(V_T^0\bigl(x,{\phi}^*\bigr)+W\bigr)\bigr]\\
&&\qquad=
E\biggl[\tilde{U}\bigl(y^* (dQ/dP)\bigr)+y^* (dQ/dP) \biggl(\int
_0^T G_t^*(Z_t -
S_t) \,dt + W \biggr) \biggr]+ y^* x.
\end{eqnarray*}
For all $\phi\in\mathcal{A}'(U,c)$
\begin{eqnarray*}
&&E\bigl[U\bigl(V_T^0(x,\phi)+W\bigr)\bigr]\\
&&\qquad\le E\biggl[
\tilde{U}\bigl(y^* (dQ/dP)\bigr)+y^* (dQ/dP) \biggl(\int_0^T
G_t^*(Z_t - S_t) \,dt + W \biggr) \biggr]+
y^* x,
\end{eqnarray*}
again by \eqref{mobel}. Hence, the strategy ${\phi}^*$ is indeed optimal.
\end{pf}

%le5.6 #&#
\begin{lemma}\label{martingale} Under the assumptions of the previous
theorem, any
$\phi\in\mathcal{A}'(U,c)$ satisfies
\[
E_Q\int_0^T \phi_t
Z_t \,dt=0.
\]
\end{lemma}

\begin{pf} Assume $T=1$. Define
\[
\Phi^+_t:=\int_0^t (
\phi_s)_+\,ds,\qquad \Phi^-_t:=\int_0^t
(\phi_s)_-\,ds.
\]
We show that $E_Q\int_0^1 Z_t \,d\Phi^+_t - E_Q\int_0^1 Z_t \,d\Phi^-_t=0$.

Since $\phi\in\mathcal{A}'(U,c)$, \eqref{luk}, \eqref{lak} and H\"
older's inequality imply that\break $E_Q[V_1^0(x,\phi)]_-<\infty$,
hence Lemma~\ref{morecamb} implies that
\[
E_Q\int_0^1 |
\phi_t|^{\beta}\bigl(1+|S_t|\bigr)^{\beta}\,dt<\infty,
\]
a fortiori,
%
%e46 #&#
\begin{equation}
\label{bbbf} E_Q\bigl(\Phi^+_1\bigr)^{\beta}=E_Q
\biggl(\int_0^1 (\phi_t)_+ \,dt
\biggr)^{\beta
}<\infty.
\end{equation}

Define $\Phi^+_t(n):=\Phi^+(k_n(t)/n)$ where
\[
k_n(t):=\max\biggl\{i\in\mathbb{N}\dvtx \frac{i}{n}\leq t
\biggr\}
\]
and observe that $\,d\Phi^+_t(n)\to \,d\Phi^+_t$ a.s. in the sense of weak
convergence of measures on $\mathcal{B}([0,1])$.
As $Z_t$ is a.s. c\`adl\`ag, its trajectories have countably many
points of discontinuity (a.s.). By $d\Phi^+_t\ll \operatorname{Leb}$,
this implies
\[
Y^+_n:=\int_0^1 Z_t
\,d\Phi^+_t(n)\to\int_0^1
Z_t \,d\Phi^+_t=:Y^+,
\]
almost surely. Furthermore,
%
%e47 #&#
\begin{eqnarray}
\label{fura}&& \biggl\llvert \int_0^1
Z_t \,d\Phi^+_t(n)\biggr\rrvert
\nonumber
\\[-8pt]
\\[-8pt]
\nonumber
&&\qquad =\Biggl\llvert \sum
_{k=1}^n Z_{k/n}\bigl[\Phi
^+_{k/n}(n)-\Phi^+_{(k-1)/n}(n)\bigr] \Biggr\rrvert
\leq\sup
_t |Z_t| \Phi^+_1,
\end{eqnarray}
where $\sup_{t\in[0,T]} |Z_t|\in L^{\gamma'}(Q)$ by assumption and
$\Phi
^+_1\in L^{\beta}(Q)$ by \eqref{bbbf}.\vspace*{1pt}
It follows by H\"older's inequality that the sequence $Y^+_n$ is
$Q$-uniformly integrable,\vspace*{1pt} so
$E_QY^+_n\to E_QY^+$, $n\to\infty$.
From \eqref{fura}, we get, noting that $\Phi^+_0(n)=0$,
%
%e48 #&#
\begin{eqnarray}
\label{eq:stasta} E_QY^+_n&=& E_Q \Biggl[\sum
_{l=0}^{n-1} (Z_{l/n}-Z_{(l+1)/n})
\Phi^+_{l/n}(n) \Biggr] + E_QZ_1
\Phi_1^+(n)
\nonumber
\\[-8pt]
\\[-8pt]
\nonumber
&=& E_Q Z_1\Phi^+_1(n),
\end{eqnarray}
by the $Q$-martingale property of $Z$. Analogously, as $n\rightarrow
\infty$,
\[
E_QY^-_n=E_Q Z_1
\Phi^-_1(n)\to E_Q Y^-,
\]
where $Y^-_n$ is defined analogously to $Y_n^+$ using $d\Phi^-_t$
instead of $d\Phi^+_t$ and
\[
Y^-:=\int_0^1 Z_t \,d
\Phi^-_t.
\]
Since $\Phi_1(n)=\Phi_1=0$, \eqref{eq:stasta} implies that
$E_Q(Y_n^+ -
Y_n^-)=0$ for all $n$, whence also
\[
E_Q\bigl(Y^+ - Y^-\bigr)=E_Q\int_0^T
\phi_t Z_t \,dt=0,
\]
completing the proof.
\end{pf}

%\begin{appendix}
%\section{}
%\end{appendix}

% zodis "Acknowledgments" paliekamas pagal autoriu
\section*{Acknowledgments}
For helpful comments, we thank Bruno Bouchard, Yan Dolinsky, Ioannis Karatzas,
Kostas Kardaras, Johannes
Muhle-Karbe, Teemu Pennanen, Walter Schachermayer,
Mete Soner, and seminar participants at the Oberwolfach
Workshop in Stochastic Analysis in Finance and Insurance.%,
%University of Vienna, and Durham University.

%\bibliographystyle{plain}
%\bibliographystyle{imsart-nameyear}
%\bibliography{ebenezer}

% imsref loaded by akundreckaite, 2014-07-17 10:11:28
%

%\begin{supplement}[id=suppA]
%\sname{Supplement A}
%\stitle{}
%\slink[doi]{10.1214/00-AAPXXXXSUPP} %[doi,text={...}] - jei reikia
%suskaldyti doi
%\sdatatype{.p.d.f.}
%\sfilename{aapXXXX\_supp.pdf}
%\sdescription{}
%\end{supplement}

%\begin{thebibliography}{99}
%\bibitem{r1}
%\bibitem{r1}
%\end{thebibliography}

\printaddresses
\end{document}